**Title**

# Verwey transition as evolution from electronic nematicity to trimerons via electron-phonon coupling

**Authors**

Wei Wang[1*], Jun Li[1#], Zhixiu Liang[1], Lijun Wu[1], Pedro M. Lozano[1,2], Alexander C. Komarek[3], Xiaozhe Shen[4], Alex H. Reid[4], Xijie Wang[4], Qiang Li[1,2], Weiguo Yin[1], Kai Sun[5], Ian K. Robinson[1,6], Yimei Zhu[1], Mark P.M. Dean[1], Jing Tao[1*, ##]

**Affiliations**

[1]Condensed Matter Physics and Materials Science Division, Brookhaven National Laboratory, Upton, NY 11973, USA.

[2]Department of Physics and Astronomy, Stony Brook University, Stony Brook, NY 11794-3800, USA.

[3]Max Planck Institute for Chemical Physics of Solids, Nöthnitzer Street 40, 01187 Dresden, Germany.

[4]SLAC National Accelerator Laboratory, Menlo Park, CA 94025, USA.

[5]Department of Physics, University of Michigan, Ann Arbor, Michigan 48109, USA.

[6]London Centre for Nanotechnology, University College, London WC1E 6BT, UK.

[#]Current address: Beijing National Laboratory for Condensed Matter Physics, Institute of Physics, Chinese Academy of Sciences, Beijing 100190, China

[##]Current address: Department of physics, University of Science and Technology of China, 230026 Hefei, Anhui, China

*Corresponding authors: wangweipositron@outlook.com; jingtao1975@gmail.com

**Abstract**

Understanding the driving mechanisms behind metal-insulator transitions (MITs) is a critical step towards controlling material's properties. Since the proposal of charge-order-induced MIT in magnetite $Fe_3O_4$ in 1939 by Verwey, the nature of the charge order and its role in the transition have remained elusive. Recently, a trimeron order was discovered in the low-temperature structure of $Fe_3O_4$; however, the expected transition entropy change in forming trimeron is greater than the observed value, which arises a reexamination of the ground state in the high-temperature phase. Here we use electron diffraction to unveil that a nematic charge order on particular Fe sites emerges in the high-temperature structure of bulk $Fe_3O_4$, and that upon cooling, a competitive intertwining of charge and lattice orders arouses the Verwey transition. Our findings discover an unconventional type of electronic nematicity in correlated materials and offer innovative insights into the transition mechanism in $Fe_3O_4$ via the electron-phonon coupling.

**Teaser**

Verwey transition could be evoked by the emergent of a nematic order interacting with specific phonon modes upon cooling.

## MAIN TEXT

### Introduction

The Verwey transition in magnetite, $Fe_3O_4$, signaled by a sudden decrease in electrical conductivity by a factor of 100 upon cooling, is known to be associated with a structural phase transition from cubic to monoclinic at a temperature around the Verwey temperature, $T_v$ = 125 K (*1–3*). As the first transition metal oxide material with proposed charge-order-induced MIT and structural transitions, searching for the origin of charge ordering and charge-lattice interactions in $Fe_3O_4$ have captivated tremendous research efforts for nearly a century (*4–6*). However, the role of charge ordering in this material has been a longstanding debate. On one hand, long-range charge order was argued to be a result, rather than a cause, at the critical point ($T_v$) of the transition (*7–9*), leaving the mechanism for the phenomenon unresolved. On the other hand, short-range lattice distortion and electronic instability have been reported in the electron, x-ray and neutron scattering studies (*10–14*), no experiments to date observe any long-range charge order above the critical temperature $T_v$. Theoretical studies have even suggested that potential long-range order would be unstable(*15*). Hence the Verwey transition mechanism is a heavily-clouded subject. The technical difficulty of searching charge order at and above $T_v$ may lie in the material's integrity including oxygen stoichiometry (*16*, *17*), distinct surface and bulk quality and properties (*18–20*).

In this work, we study the symmetry breaking on electronic structure and explore the interaction between electrons and lattice in $Fe_3O_4$ using electron scattering. The extra reflections observed in the diffraction pattern indicate a long-range charge order, i.e., nematic phase, embedded in the high-temperature phase. Using femtosecond laser pulses, we observed a strong electron-phonon coupling in the transient state with MeV electron probe. We provide a linkage between the nematic order and trimeron order to advance the understanding of Verwey transition.

## Results

**Electron diffraction results using a transmission electron microscope (TEM).** The TEM experiment was performed on two single crystals with slightly different stoichiometries (Sample 1 $Fe_{3(1+0.0005)}O_4$ and Sample 2 $Fe_{3(1-0.001)}O_4$, see electronic resistance measurement in Fig. S1 and the Supplementary Materials for additional details). The diffraction patterns as a function of temperature from Sample 1 and 2 are presented in Figs. S2-3. Both samples present an abrupt increase in the electronic resistance measurement at $T_v$. At low temperatures, the superlattice reflections corresponding to the superstructure with trimerons are observed in both samples in the diffraction patterns. The slightly different transition temperatures are due to the different stoichiometries of the samples, and all these features indicate that Sample 1 and 2 are high-quality crystals and qualitatively the same, as shown in Fig. 1b. For simplicity, all the reflections and orientations in this paper will be defined with reference to the cubic structure. It is very surprising that the {200} reflections with weak intensity in the <001> zone electron diffraction pattern, which should be forbidden for the 16*d* site of Fe ions in the cubic structure (Space group: $Fd\bar{3}m$, No.227) (*21*, *22*), are clearly present over a wide range of temperatures above $T_v$. The "forbidden" reflections are sharp, with similar widths to the main Bragg peaks, i.e., a long-range order associated with the "forbidden" reflections is likely present. We have examined many sample flakes with various sample thickness in the TEM, and we find that the {200} reflections are undetectable in the very thin areas but statistically dominant in the relatively thick areas (typically tens of nanometers in thickness). This indicates that the "forbidden" reflections are not from the sample surface due to the surface contamination or reconstruction but are intrinsic to the bulk. We also note that these 'forbidden" reflections at [001] zone axis cannot be introduced by multiple scattering since the lack of a combination of any crystallographically allowed fundamental Bragg reflections to generate {200} reflections in the reciprocal space. To verify this point, the dynamic diffraction simulation was performed by considering the appropriate sample thickness and multiple scattering effects, and the corresponding simulation result is shown in Fig. S4.

**The origin of the "forbidden" reflections.** Firstly, since more reflections, e.g., {100} and {110}, would present in the case (see Fig. 2a) of the monoclinic phase (Space group: *Cc*, No.9) which occurs below $T_v$, the appearance of only {200} reflections in Fig. 1b rules out the existence of a lower symmetry with the monoclinic structure embedded in the high-temperature phase. Secondly, we found that the {200} reflections could present when we assign different valence charges at octahedral Fe sites, which are widely considered as the candidate ions that induce the potential charge ordering (*3*). According to the $Fd\bar{3}m$ symmetry, we can preserve the three-fold rotational symmetry along one of <111> families of directions as shown in Fig. 1d, while breaking the other three three-fold rotational symmetries by assigning different valence charge at 16*d* Fe ions, as shown in Fig.1e and Fig. S5. This type of charge ordering converts the $Fd\bar{3}m$ space group to $R\bar{3}m$ space group and gives rise to {200} reflections in the diffraction simulations in Fig. 1f. According to the charge arrangement, there are four twin variants in the charge ordering models (see the Supplementary Materials for additional details and Fig. S5). Dynamic electron diffraction simulations based on models with adjustable valence charge values were carried out and compared with experiments (Fig. 1g and Fig. S6), revealing the charge differentiation between the two ordered Fe ions could be 0.05 $e^-$.

We note that atomic displacements of octahedral Fe atoms with the same symmetry breaking indicated in Figs. 1d and 1e could also give rise to the "forbidden" reflections in diffraction patterns. Our dynamic electron diffraction simulations and the comparison with the experimental data reveal possible existence of a pure atomic displacement wave with an amplitude of ~0.05 Å (For the dynamic electron diffraction calculation on atomic displacement, see the Supplementary Materials for additional details and Fig. S7). However, in that the case, neutron diffraction would be sufficient to detect it with much better accuracy in ascertaining atomic positions and lattice symmetry by neutron diffraction techniques. Indeed, $Fe_3O_4$ crystals are known to have diffuse scatterings in proximity to the Bragg peaks at the temperature just above $T_v$ and the intensities of those diffuse scatterings are comparable to the "forbidden" {200} reflections in electron diffraction (The intensity comparison of diffuse scattering and {200} peaks are shown in Fig. S8). Neutron diffraction results have clearly reported the observations of the diffuse scattering but did not detect the {200} reflections at temperatures above $T_v$ (*7*, *10*, *11*). These facts strongly suggest that, although we cannot rule out the possibility of lattice symmetry breaking, the intensity of the "forbidden" {200} reflections must mainly come from an electronic ordering. To quantitatively separate the electron and lattice contribution to {200} reflections, more experiments are desirable. Applying external anisotropic strain to tune the lattice structure can be insightful explore the competition between electronic and lattice structures (*23*, *24*). Additionally, the intrinsic electronic response to the lattice distortion can also be extracted via measuring the elastic shear modulus as shown in (*25*). Moreover, the inelastic light scattering is a beneficial tool to identify the lattice structure and electronic nature simultaneously cross the Verwey transition (*26*).

In a prior resonant x-ray diffraction study, the observation of {200} reflections was ascribed to anomalous tensor scattering from the anisotropy of the local electron density on the 16*d* Fe ions (*27*). This occurs because the orbital orientation at the 16*d* site follows the $\bar{3}m$ crystallographic point group symmetry at these sites. For the electron scattering cross section, {200} reflections are extinct since the $Fd\bar{3}m$ space group symmetry is preserved. (The reflection conditions for $Fd\bar{3}m$ space group symmetry, *h*00: *h* = 4*n*; 0*k*0: *k* = 4*n*; 00*l*: *l* = 4*n*, *n* = integers (*28*). Thus, the anomalous scattering case discussed in (*27*) cannot explain what we observed in the electron diffraction data. In our charge ordering model as shown in Figs. 1d and 1e, the contribution of 16*d* Fe ions to {200} reflections are no longer canceled out due to the different charge among 16*d* Fe ions, which is the best solution in our study. Moreover, electron scattering is more sensitive to valence charge ordering than x-ray techniques due to its strong scattering power and nature of interaction with electrostatic potentials in the sample (*29*). To exam the different sensitivities to Fe ions, we calculated the electron and x-ray form factors for $Fe^{2+}$ and $Fe^{3+}$ ions, as shown in Fig. S9. The calculated result illustrates that, comparing with x-ray scattering, electron scattering is more sensitive to the disproportion of the valence electrons at low scattering angles, e.g., {200} reflections, which may explain why those “forbidden” reflections can be observed using electron diffraction. Furthermore, although the resonant x-ray scattering enhances the sensitivity to the electronic states(*30*), the small charge disproportion, i.e., ~ 0.05 $e^-$, may be too weak to be detected using the resonant x-ray anisotropic tensor susceptibility mechanism in $Fe_3O_4$, as suggested in (*31*, *32*). Therefore, we conclude that the different sensitivities of the x-rays and electrons to the different types of electronic structures lead to a comprehensive study on electronic state in $Fe_3O_4$, i.e., anisotropic electron density state with different probabilities on 16*d* Fe ions.

**Electronic nematic phase above $T_v$.** Since the amount of the long-range ordered lattice distortions here appear to be below the sensitivity of standard neutron diffraction experiments and the symmetry of the electron distribution is unambiguously broken from the cubic phase, the anisotropic electronic structure in $Fe_3O_4$ presents a paradigm of an electronic liquid crystal (ELC) phase at the temperature above $T_v$. The ELC phases are quantum states of matter where strong interactions between electrons drive an instability that spontaneously breaks certain point group and space group symmetry of a crystal (*33*), i.e., describing a state with a modulated electronic structure. Among various electronic liquid crystal phases, we found that the most relevant one for our study is the electronic nematic phase, where certain rotational symmetry is spontaneously broken, while the translation and space inversion symmetries are preserved. As we mentioned above, in the charge ordering model for $Fe_3O_4$, each position on 16*d* sites has a face-centered translational symmetry, which is preserved. However, the <111>-direction three-fold rotational symmetries are broken as shown in Figs. 1d and 1e.

ELC nematic phases have been widely known for their significance in distinguishing electronic structures and lattice from symmetry perspective with electron driven mechanism, hence providing a unique way of probing the structure-property relationships in correlated materials (*34*). To date, ELC nematic phases are observed in cuprates (*35*), iron pnictides, manganites, and ruthenate, etc. (*33*), and those findings have generated far-reaching impacts in the research area. In report (*36*), the breaking of rotational symmetry was detected in copper oxide superconductors $La_{2-x}Sr_xCuO_4$ via in-plane anisotropic electron transport. And the direction of the electronic nematic order is detached from the lattice symmetry, which indicates the nematicity is not reduced by lattice distortion. All the observations are very similar to ours. However, unlike most discovered ELC phases with rotational symmetry breaking from four-fold ($C_4$) to two-fold ($C_2$), the present ELC phase in magnetite breaks the $O_h$ point group down to $D_{3d}$, which breaks a different set of three-fold rotational symmetry (Detailed analysis about the point group on the charge ordering model see Supplementary Materials). After revisiting the original definition of nematic phases in classical liquid crystals, we conclude that the charge order in $Fe_3O_4$ is an unconventional type of electronic nematicity ever reported in correlated materials based on the following analysis. In classic liquid crystals, the disordered phase preserves the continuous 3D rotational symmetry [SO(3)] and nematicity is characterized by a director (unit vector) order parameter, i.e., a traceless symmetric matrix with five independent components (*37, 38*). In a cubic lattice with $O_h$ point group symmetry, this 5-dimensional representation splits into a two-dimensional $E_g$ representation and a three-dimensional $T_{2g}$ representation. The two components in the $E_g$ representation characterize the $C_4$ to $C_2$ symmetry breaking, matching the observations in all other reported cases. In contrast, the $T_{2g}$ representation with three components that breaks the $O_h$ symmetry down to $D_{3d}$ are the nematic order parameters, i.e., electronic order parameter, observed here for $Fe_3O_4$, which notably broadens the existence of electronic nematicity in electron correlated materials and provides an advanced pathway to study the entanglement between the electrons and the lattice.

It is worthwhile to highlight this nematic signature is observed above the metal-insulator transition temperature ($T > T_v$), where charge carriers remain mobile. Since conventional charge order is expected to be washed away by mobile charge carriers, this observation suggests that the electronic nematicity plays a crucial role in the high-temperature phase. This is consistent with neutron scattering results (*7*, *10*, *11*), which does not observe lattice distortion or any building blocks of directional nature.

**{200} reflection intensity evolution upon cooling and warming.** To further demonstrate the role of this charge order, or electronic nematicity, in the Verwey transition, we measured the intensity variation as a function of temperature during *in situ* cooling and heating experiments on $Fe_3O_4$ single crystals. Typical electron diffraction patterns above and below $T_v$ are shown in Fig. 2a. The {200} reflections are denoted as the charge ordered (CO) peaks at 300 K, while the SL reflections, i.e., "forbidden" reflections, are consistent with the monoclinic structure with trimerons below $T_v$. Figure 2b are the intensity measurements of the CO and SL peaks, and the measurements from the two samples are very similar. The CO peak intensity reflects the amplitude of the nematic order parameter. (The intensity of CO reflection as a function of valence charge on octahedral Fe sites is calculated and shown in Supplementary Materials in Fig. S10, the larger charge disproportion, the higher peak intensity.) The nematic signature remains nonzero above $T_v$, even up to room temperature. As the temperature decreases towards $T_v$, a rapid increasing of the nematic order parameter is observed below 170 K. This temperature-dependence invites comparisons to the nematic order observed in the high-temperature superconductor $YBa_2Cu_3O_{6+x}$ (*33*, *39*).

Additionally, peak width as a function of temperature measured from two samples is shown in Fig. 2c. The line profiles at a few typical temperatures are presented in Fig. 2d. The sharp CO reflections at all the temperatures demonstrate that the charge ordering phase is a long-range order. According to our measurements of the peak width, the coherent length of the nematic order is about 7.8 nm at 295 K and gradually increases to a value about 10.7 nm at 157 K in Sample 1. Sample 2 displays a similar temperature evolution with slightly larger coherent lengths upon cooling. Right at $T_v$, strong SL reflections appear and their intensity increases as the temperature cools down. Note that the CO peaks are coincided with in the SL reflections in the low-temperature phase; thus it is difficult to separate them in static observations. The results manifest interplay of the electronic nematicity and the lattice order through thermal processes, which on one hand confirms the mechanism of charge-order-induced transition in this material. When the sample was heated up from room temperature to higher temperatures, the {200} peak intensity got reduced and the peak width got broader. Since the sample was damaged above 423 K, we deduce that the hypothetical temperature at which the nematicity disappears is higher than 423 K and does not necessarily exist at all. The electron diffraction patterns taken during the heating process and the corresponding measurement results are shown in Supplementary Materials in Fig. S11.

The evolution from electronic nematicity to trimerons is a topic of interest in condensed matter physics because it provides insights into the underlying mechanisms that drive the formation of complex electronic phases in materials. For example, electronic nematicity (rhombohedral; cubic symmetry is already broken) can help to stabilize the trimeron order (with a monoclinic supercell) by providing a preferred directionality for the electronic degrees of freedom that participate in trimeron formation. Furthermore, the discovery of the electronic nematicity reveals that the Verwey transition belongs to a new universality class based on theoretical studies of ELC phase transitions, different from the conventional picture (See details in section "Two possible scenarios in the Verwey transition upon cooling" in Supplementary Materials).

**Transition from electronic nematicity to trimerons.** We find interesting relationships that bridge the charge arrangements in electronic nematicity above $T_v$ and the trimeron order below $T_v$ via electron-phonon coupling. Since four variants of electronic nematicity exist equally, there can be twin structures with an order as spatial distribution, or a random mixture along the ***c*** axis in a local volume. Note that two phonon modes, $\Delta_5$ and $X_3$ with their wave vectors $\mathbf{k}_\Delta$ = (0, 0, 1/2) and $\mathbf{k}_X$ = (0, 0, 1) in cubic notation, respectively, were considered to play substantial roles in the structural phase transition at $T_v$ (*7*, *10*, *40*, *41*). Since the ***c*** axis is doubled from cubic phase to monoclinic phase, the $\Delta_5$ mode with a (0, 0, 1/2) wave vector may help to regulate the stacking order of electronic nematicity along the ***c*** direction during the formation of trimeron order. Meanwhile, the $X_3$ mode (together with other possible phonon modes, such as polaron tunneling (*42*, *43*)) in its wavelength may destroy the sheet-like charge ordering in electronic nematicity and create the linkage of charged atoms between the sheets to be trimeron orders. The atomic layers of octahedral Fe atoms with different valence in unit-cell configuration of the cubic and monoclinic phases are shown in Figs. S13 and S14. The present findings bestow more character to the $\Delta_5$ and $X_3$ phonon modes, which were initially selected based on the symmetry breaking elements in *lattice* through the transition (*40*, *41*), suggesting they could be essential to transform a nematic phase in *electron* substructure into the 3D trimeron order. To figure out how these phonon modes assist the dimensional switch of the electronic order during Verwey transition, MeV ultrafast electron diffraction (UED) method was employed to detect the dynamic behavior of electron and phonon modes in the far-from-equilibrium state.

**UED study on the electron-phonon coupling in $Fe_3O_4$.** Direct observations of the coupling between charge and specific phonon modes can be obtained from MeV UED results at the temperature ~ 30 K where the structure is monoclinic with the trimeron charge ordering. In the optical conductivity measurements, three main features at 0.6, 2.0 and 5 eV are related to intersite transitions between octahedral-site Fe ions, octahedral and tetrahedral-sites Fe ions, and charge transfer between O 2*p* and Fe 3*d* states, respectively, which has been confirmed by density functional theory calculations (*44*–*46*). In our UED experiment, the 1.55 eV laser pulses are employed to tune the charges between the octahedral-sites Fe ions in trimerons, which quenches the ordering phase, excites the structural lattice, and partially drives the system from the low-temperature phase to high-temperature phase, as suggested in (*43*, *47*, *48*). Figure 3a shows the experimental setup of the excitation and observation processes, i.e., a pump-probe scheme. Since the Bragg peaks are more sensitive to the lattice structure, we prepared a UED single-crystal sample at the [110] zone axis for intensity measurements from more Bragg / SL peaks than the [001] zone shown in Fig. 1 and Fig. 2. The intensity variation as a function of delay time is plotted in Fig. 3b which gives two timescales during dynamic processes. The fast one takes place in the first 700 fs in which all the SL peak intensities drop quickly while no obvious change can be measured from Bragg peak intensities. A slow dynamic process follows, which starts from ~ 700 fs and lasts at least a few ps in our observations, which is consistent with the timescale observed in the ultrafast x-ray diffraction study (*47*). During this period, only minor changes were observed from SL peaks while both increase and decrease in intensities were observed from Bragg peaks. Additional intensity measurements for the Bragg peaks can be found in Supplementary Materials. From the intensity variation in the time domain as shown in Fig. 3b and Fig. S16, we found that there is no observable oscillation behavior, implying that the photoexcited states are not induced by any coherent phonon modes (*49*, *50*). The absence of coherent modes may rely on multiple factors, e.g., pump energy, pump fluence, probe energy, etc. (*43*, *45*).

To understand the origin of the peak intensity variations, we performed dynamic electron diffraction analysis considering atomic charge distribution, atomic displacements, and existence of twin variants (see Tables S4 and S5 about the twin variants and corresponding reflections in Supplementary Materials). We calculated intensities of Bragg peaks and SL peaks by changing the valence electron on the octahedral sites in the monoclinic phase in Fig. S17. The calculation indicates that the SL reflections are much more sensitive to the charge states while the Bragg peaks are more sensitive to the atomic displacements (see Figs. S19, S20). The diffraction calculations reveal that the intensity change of the SL peaks mainly comes from the change in electronic order parameter and the intensity variations of the Bragg peaks can be mostly attributed to atomic displacement, i.e., lattice distortions. As illustrated in Fig. 3c, during the delay time range in our experiments, electrons are excited and thermalized to reduce the electronic order parameter in trimerons in the first 700 fs after photoexcitation.

In the following a few ps, lattice responds with atomic displacements from the initial monoclinic structure. In this step, we tested the effects of atomic displacements based on the $\Delta_5$, $X_3$ and a few other frozen phonon modes that researchers often consider in this material (*51*, *52*), and find that atomic displacements based on only the $X_3$ phonon mode (Fig. 3c) can provide consistent intensity variation with the experimental results (The intensity variations induced by the atomic displacements following different phonon modes are discussed in Methods and Supplementary Materials). Thus, we conclude that, during the course of photoexcitation and relaxation that is different from the thermal process, the $X_3$ phonon mode was preferably excited after the electron excitation at least with much more intensity responses than other phonon modes reflected in our UED measurement. As we mentioned above, the intensity variation without the coherent oscillation indicates that the photoexcited $X_3$ phonon mode revealed using our experimental condition is likely not a coherent phonon and suggests that the change in the electron-electron interaction modifies the potential well of the $X_3$ phonon mode during the photoexcitation process as mentioned in (*53*). Namely, the coupling between electrons and $X_3$ phonon modes is stronger than other electron-phonon couplings in $Fe_3O_4$ crystals. This finding implies that $X_3$ phonon modes play a major role in the nematic phase - trimeron phase transformation, which is responsible for regulating the charge transfer among the Fe ions in the layers and among layers. Moreover, first-principles calculations indicate that the $X_3$ phonon mode is also in charge of the electric properties in $Fe_3O_4$ – being an insulator by opening a gap at the Fermi surface during the phase transition (*9*, *41*, *54*).

What's more, the mechanisms of Verwey transition have been enormously debated for decades (*55*). The electron, phonon degrees of freedom and their interaction are the central topic, e.g., the electron-phonon coupling (*53*), order-disorder phase transition (*56*), electron-electron interaction (*57*) are proposed to be crucial for Verwey transition. To figure out the transition mechanisms, various ultrafast methods have been used to explore dynamic behaviors of electronic structure and lattice structure. For example, using the time-resolved optical reflectivity measurements, the metallic and insulating phase separation was observed during a specific pump fluence regime in 1.55 eV laser pulses as shown in (*47*, *48*). In (*45*), it was shown that the ultrashort laser pulses with 3.1 eV generated symmetry-forbidden phonon modes, which is related to the critical fluctuations of electronic order due to the electron-phonon couplings. In our UED observations, we focused on the electron-phonon interactions and provided more direct evidence by taking advantage of electron sensitivities to both valence charge and lattice (*58*). We found that the $X_3$ phonon mode was pinpointed during the photoexcitation process through the electron-phonon coupling, which was proposed in the theoretical studies (*40*, *41*). We conclude that the intimate coupling between charge and the phonon mode plays a vital role in the Verwey transition, which supports the mechanism of electron-phonon coupling in the debate. Last but not least, the $X_3$ phonon excitation in dynamic observations may imply a manipulation of the electric properties of $Fe_3O_4$ in a short time range of a few ps.

## Discussion

In summary, the discovery of the surprising symmetry breaking in the electronic state of $Fe_3O_4$ crystals at the temperature range far above $T_v$ opens a window to the further works, including theoretical interpretation of the nature of the unconventional type of electronic nematicity and its role in the correlated materials. Additionally, the findings here advance the understanding of the Verwey transition: it suggests that the high-temperature phase charge ordering phase contributes to the phase transition. In the absence of the additional measurements, e.g., applying strain control, it is not possible to explicitly rule out the lattice distortion contribution to the symmetry breaking, although all the existing measurements indicate a role for electronic nematic order at the temperatures above $T_v$. The findings described in this report connect the electronic nematicity order in the cubic lattice to the trimeron order in the monoclinic lattice through electron-phonon coupling, offering a transformative interpretation of the Verwey transition mechanism and innovative insights into the origin of the trimerons formation. The MeV UED pump-probe experiments not only provide a means to directly observe electron-phonon coupling through the dynamic process, but also suggest a possible manipulation of the materials' electric property by a specific phonon excitation using ultrafast laser pulses.

## Materials and Methods

### Sample preparation

We chose $Fe_3O_4$ bulk single crystals from two resources, one is a mineral procured from SurfaceNet GmbH (Sample 1) and the other is a laboratory synthesized crystal, which was grown at the Max-Planck Institute for chemical physics of solids in Dresden (Sample 2). Resistance measurements show that both crystals have a first-order MIT, and the transition temperature $T_v \sim 115$ K for Sample 1 and $T_v \sim 123.4$ K for Sample 2 (see Supplementary Materials). The UED sample was cut from Sample 1 and prepared using mechanical polishing and Focused Ion Beam and thinned down to electron transparent.

### MeV ultrafast electron diffraction

The experiments were performed on the MeV-UED (*59*, *60*) setup at Stanford Linear Accelerator Laboratory which can achieve < 150 fs (FWHM) temporal resolution. Using this experimental setup, the sample was excited by a 75 fs (FWHM) laser pulse with a photon energy of 1.55 eV and a fluence of 4 mJ/cm$^2$. The optical pulses at a repetition rate of 180 Hz were focused on the sample. The laser pump pulse and probing electron beam are colinear in the UED instrument. The sample were excited by synchronized 3.5 MeV electron pulses containing ~$10^6$ electrons with an illuminated area of 100 µm in diameter. The sample temperature of 30 K was achieved using a conducting sample holder by liquid helium.

### KeV transmission electron diffraction

The diffraction patterns shown in Figs. 1b and 2a were obtained from a JEOL ARM 200F microscope operating at 200 kV at Brookhaven National Laboratory. *In situ* cooling TEM experiment was carried out using the Gatan liquid nitrogen cooling stage with double tilt capability. The experiments were carried out at the lowest attenable temperature ~ 90 K. Two methods were used to prepare the samples: 1) crushing the sample into pieces and suspended on the TEM Cu grid coated with carbon thin film; and 2) using Focused Ion beam.

### Electron diffraction simulations for TEM and UED data

To understand the origin of the intensity variation in the electron diffraction patterns, we performed dynamic diffraction simulation based on the Bloch wave method using the computer code developed in-house. See Supplementary Materials for the dynamic simulation.

## Acknowledgments:

**Funding:** Work at BNL was supported by US Department of Energy (DOE), Office of Science, Office of Basic Energy Sciences (BES), Materials Sciences and Engineering Division under Contract No. DE-SC0012704. The experimental MeV-UED part of this research was performed at the SLAC MeV-UED facility, which is supported by the U.S. Department of Energy BES SUF Division Accelerator and Detector R&D program, the LCLS Facility, and SLAC under Awards No. DE-AC02-05-CH11231 and No. DEAC02-76SF00515. Work performed at UCL was supported by EPSRC. For the purpose of open access, the author has applied a Creative Commons Attribution (CC BY) licence to any Author Accepted Manuscript version arising. This work was also supported by the resources of Center for Function Nanomaterials (CFN) at BNL.

**Author contributions:** J.T. and J.L. conceived the experiment. J.T., W.W., Z.L. collected the TEM and MeV-UED data. A.C.K. grew the single crystal for the TEM experiment. W.W. and Z.L. prepared the samples for TEM and UED study. L.W. developed the in-house code for the diffraction simulation. P.M.L. and Q.L. preformed the electronic resistance measurements. SLAC MeV-UED team consisting of X.S., A.H.R., X.W. assisted the experiments. J.T., W.W., Z.L. and L.W. performed data analysis of the experimental TEM and UED data and the diffraction calculations. K.S. contributed to the theoretical input. J.T., W.W., L.W., W.Y., I.K.R., Y.Z., and M.P.M.D. discussed and interpreted the data. J.T. and W.W. wrote the paper with critical input from M.P.M.D., I.K.R., W.Y., Y.Z. and all other authors.

**Competing interests:** The authors declare no competing interests.

**Data and materials availability:** All data needed to evaluate the conclusions of the paper are present in the paper and/or the Supplementary Materials.

## Table of contents for Supplementary Materials

## Figures and Tables

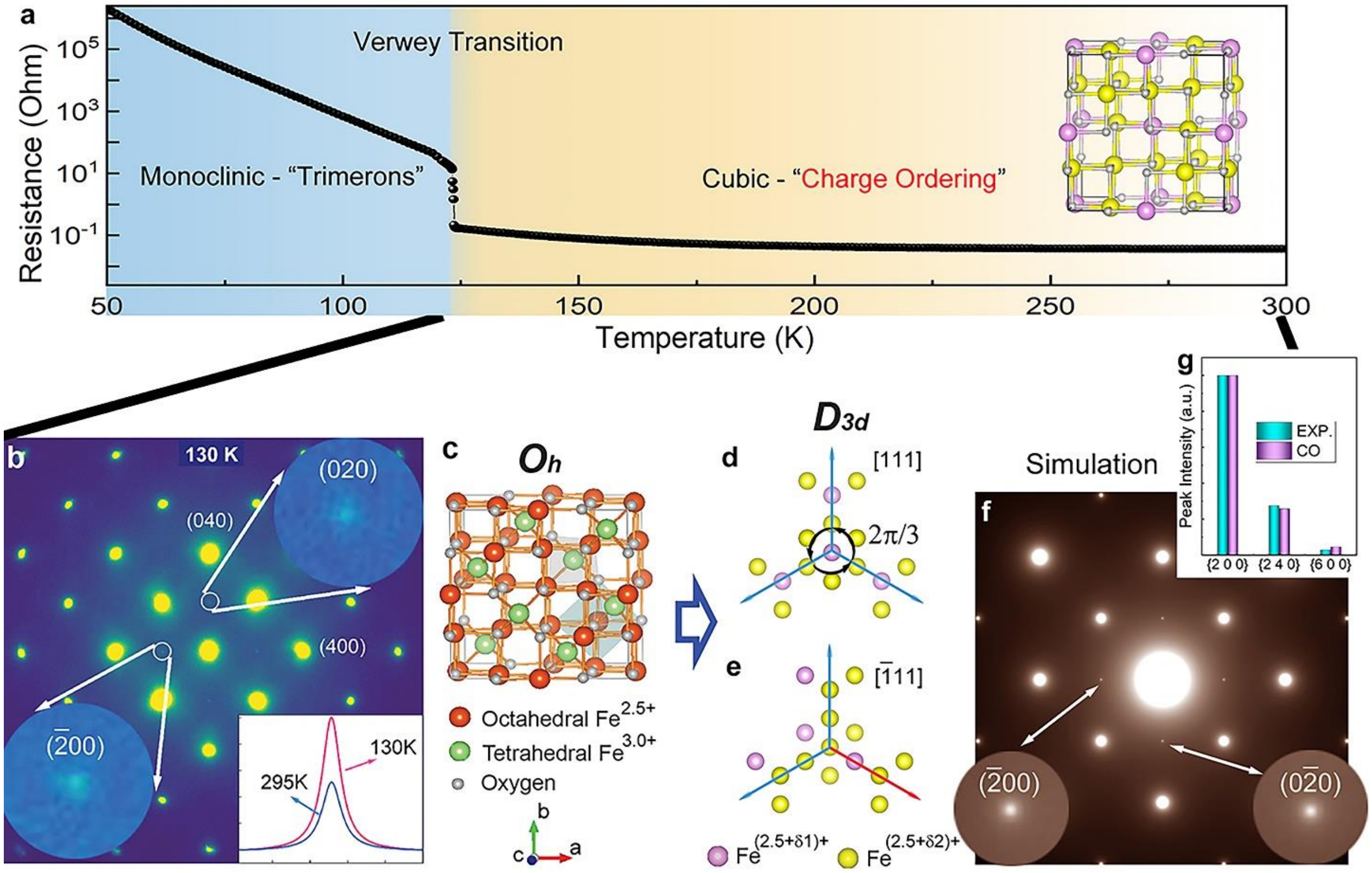

**Figure 1. A discovery of charge ordering in $Fe_3O_4$ crystals at the temperatures above the Verwey temperature. a,** Resistance measurements from single-crystal $Fe_3O_4$ shows a sharp transition at $T_v$ ~ 123 K, measured from Sample 2. Inset is a charge ordering model in the cubic phase. **b,** Upon cooling, {200} reflections, which are forbidden in the cubic structure, can be seen in electron diffraction pattern, taken from Sample 1. The inset shows the intensity profiles of the “forbidden” {200} reflections at 130 K and 295 K. **c,** A crystal model of the cubic $Fe_3O_4$, $O_h$ is the point group for the cubic structure. **d and e,** Charge distribution on the octahedral Fe sites with different valence charge values, i.e., $Fe^{(2.5+\delta1)+}$ and $Fe^{(2.5+\delta2)+}$ along [111] and [$\bar{1}$11] direction. The three-fold rotational symmetry is preserved in [111] direction shown in **d,** it is broken in [$\bar{1}$11]direction in **e**. $D_{3d}$ is the corresponding point group considering rotational symmetry breaking in the charge ordering model. **f,** Simulation of electron diffraction pattern based on the charge ordering model are shown in **d** and **e**. **g,** Normalized intensities of experimentally measured {200}, {240} and {600} reflections are compared to the results of simulations based on the charge ordering model using dynamic electron diffraction methods, showing a consistency. EXP. is the experimental data; CO is the calculation result.

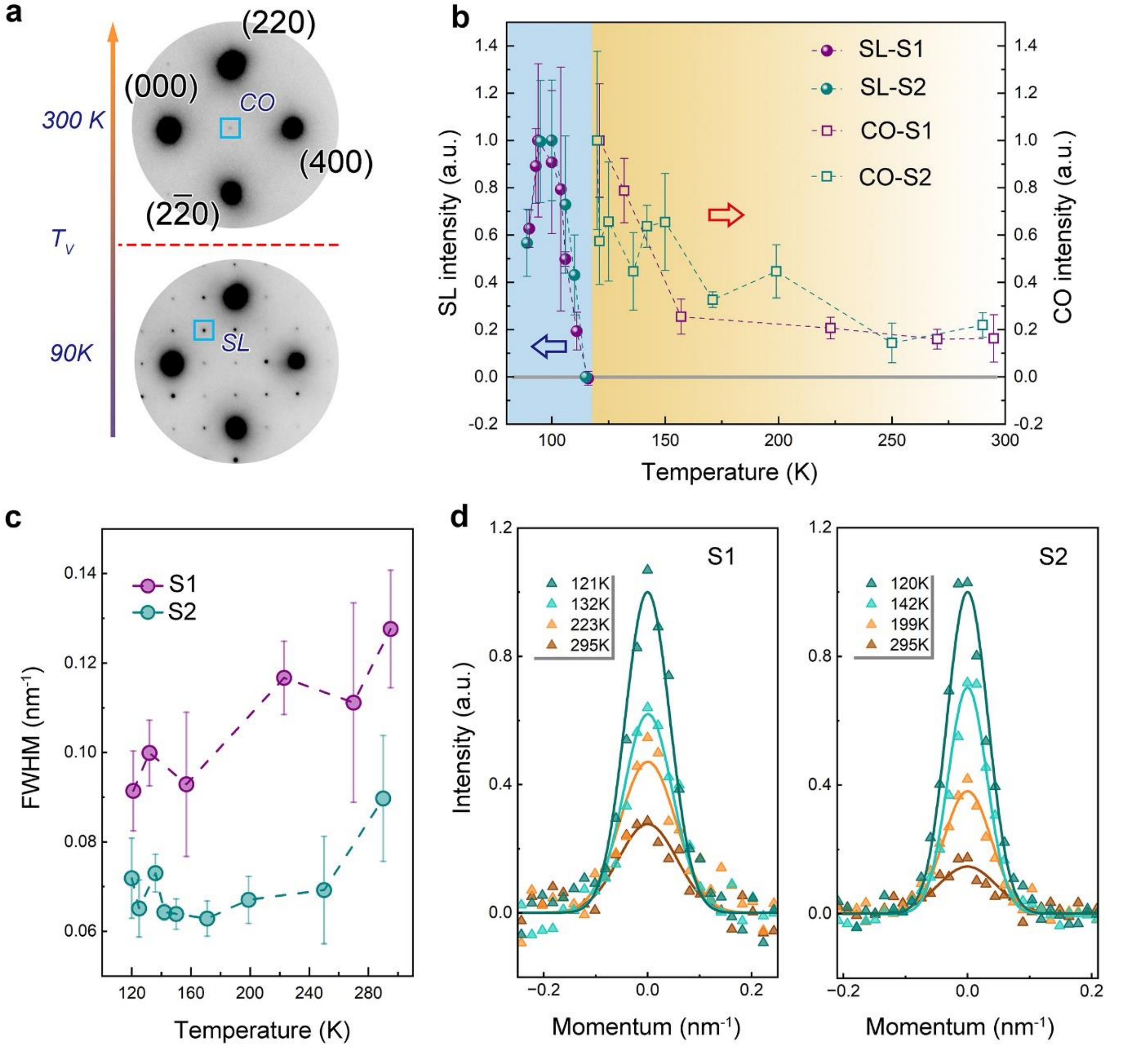

**Figure 2. Evolution of charge order and lattice order measured by reflection intensities in electron diffraction patterns upon cooling in $Fe_3O_4$. a,** Electron diffraction patterns above and below Verwey temperature $T_V$, with the appearance of the charge ordered (CO) reflection in electronic nematic phase and superlattice (SL) reflection arising from the monoclinic structure with the trimeron charge order, respectively. **b,** Intensity variations of CO and SL reflections are plotted as a function of temperature. Results were plotted from diffraction data obtained from two similar samples, Sample 1 (S1) and Sample 2 (S2). Error bars represent the standard deviation in the mean of intensities from the equivalent reflections. **c,** The variation of peak width, i.e., full width at half maximum (FWHM), of CO reflections measured from two samples upon cooling. The error bars of peak width plots were estimated based on measurement uncertainties and the instrument broadening due to diffraction limits. **d,** Line profile of CO reflection at different temperatures. The experiment data are shown as triangle symbols and the solid lines are the fitted results.

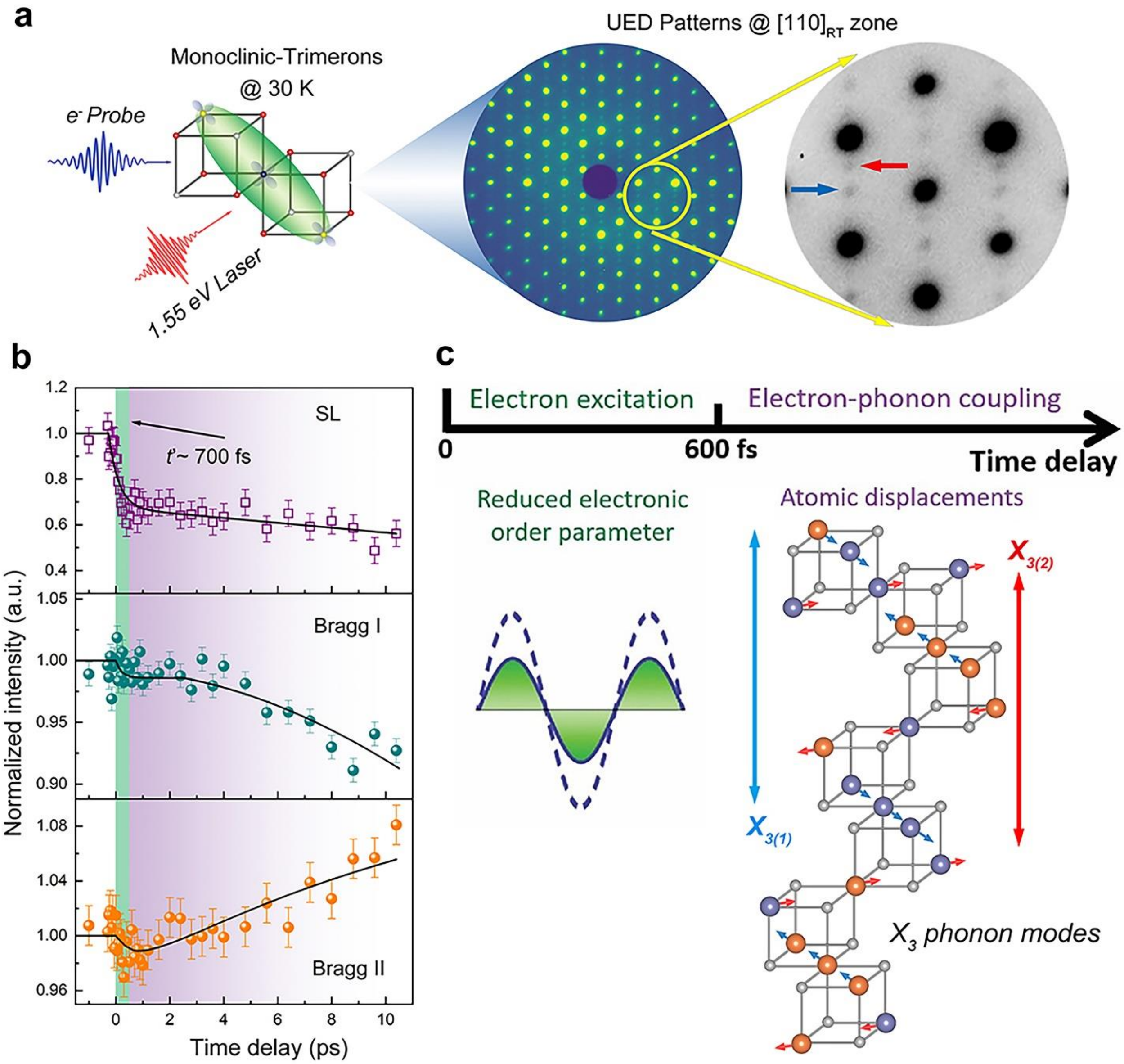

**Figure 3. Ultrafast observations of dynamic processes in photoexcited $Fe_3O_4$. a,** Schematic diagram showing 1.55 eV laser as a pump and MeV electron beam as a probe. The UED sample is from Sample 1. The UED patterns were obtained at $T$ ~ 30 K along the [110] zone axis. Arrows in the enlarged pattern indicate the SL reflections from the monoclinic structure with the trimeron order. **b,** Intensity variations of SL peaks and Bragg peaks as a function of delay time. Measurements from all the Bragg peaks show two distinct behaviors: one with intensity increase and others with intensity decrease as time goes by in a few ps. Error bars represent the standard deviation in the mean of intensity before time-zero. The solid line is a guide to the eye. **c,** Two-time regimes of the dynamic process was observed. One is the first 700 fs, labeled using t', in which the electron excitation and thermalization are dominant, resulting in a reduced charge order parameter in the system. In the subsequent second time regime, lattice responses to the reduction of electronic order parameter by atomic displacement. Specifically, analysis indicates that atomic displacements

# Supplementary Materials for

## Verwey transition as evolution from electronic nematicity to trimerons via electron-phonon coupling

Wei Wang[1*], Jun Li[1#], Zhixiu Liang[1], Lijun Wu[1], Pedro M. Lozano[1,2], Alexander C. Komarek[3], Xiaozhe Shen[4], Alex H. Reid[4], Xijie Wang[4], Qiang Li[1,2], Weiguo Yin[1], Kai Sun[5], Ian K. Robinson[1,6], Yimei Zhu[1], Mark P.M. Dean[1], Jing Tao[1*, ##]

*Corresponding authors: wangweipositron@outlook.com; jingtao1975@gmail.com

**This PDF file includes:**

Supplementary Text
Figs. S1 to S20
Tables S1 to S5

## Supplementary Text

Electronic resistance measurement results

Since the Verwey transition is sensitive to the stoichiometry of the magnetite crystal, we chose two single crystals, i.e., Sample 1 and Sample 2, to perform the electron diffraction experiment. The resistance measurement result for these two single crystals is shown below, which demonstrate the high quality of the single crystals. According to the phase transition temperature measured from Fig. S1, we estimated the stoichiometries of these two crystals based on the measurement results in(*16*): Sample 1 $Fe_{3(1+0.0005)}O_4$ and Sample 2 $Fe_{3(1-0.001)}O_4$.

Electron diffraction patterns as a function of temperature

To examine the Verwey transition in the Sample 1 (commercial) and Sample 2 (lab grown), we cooled down the samples from room temperature (300 K) to 90 K, then started to warm up and the electron diffraction patterns were captured at different temperatures in the TEM, the results are shown in Figs. S2 and S3. When the sample is transformed from spinel cubic phase to monoclinic phase, the forbidden reflections in the high-temperature cubic phase appear at low temperature due to the symmetry breaking as shown in Fig. S2b. Most of the forbidden peaks disappear above $T_v$, e.g., shown in Fig. S2c, but a few reflections, e.g., {200} reflections enlarged in inserts, with low $\boldsymbol{q}$ values are remained up to high temperature, like at 423 K in Fig. S2f. Sample 2 shows the same feature and the *in situ* cooling TEM results are in Fig. S3.

Multiple scattering effect to the {200} reflections

To verify the formation of {200} reflections in the <001> direction are not induced by multiple scattering effect, we performed the dynamic electron diffraction simulation including an appropriate sample thickness and multiple scattering effect. The sample thickness is 60 nm used in the simulation, which will be further discussed in the next section. The dynamic diffraction simulation method is based on the Bloch wave method (*61*). According to the published standard crystal structure of cubic phase (*22*), the simulated result is shown in Fig. S4. The {200} reflections are still forbidden, indicating that the {200} reflections cannot arise from multiple scattering effect in the <001> orientated diffraction pattern.

Structure factors of {200} reflections in charge ordering models

The forbidden peaks observed in the experimental pattern indicate that a symmetry breaking exists in the cubic phase. To figure out the origin of the {200} reflection, we first consider the charge ordering case. The structure factor (*F*) for each reflection (*hkl*) in the diffraction is:

$$F(hkl) = \sum_{j=1}^{N} f_j \cdot \exp\,[2\pi i(hx_j + ky_j + lz_j)]$$

Where $f_j$ is the atom form factor, $x_j$, $y_j$, $z_j$ are the atomic coordinates in the unit cell, respectively. The Wyckoff position for each element in $Fe_3O_4$ in the cubic phase is shown in Table S1.

We presume that the {200} peaks are related to the Fe ions on 16*d* symmetry sites and there are only two types of valence state on Fe ions, i.e., $Fe^{(2.5+\delta1)+}$ and $Fe^{(2.5+\delta2)+}$. Additionally, considering the absence of other "forbidden" peaks shown in the monoclinic phase, we should preserve the face centered translational symmetry in the charge ordering model. Based on these conditions, we separate Fe ions on the octahedral sites into 4 subgroups with respect to valence state on Fe ions. Namely, the atom form factor of Fe ions in each subgroup would be $f_i$, $i$ =1, 2, 3, 4.

According to the f.c.c symmetry in $Fd\overline{3}m$ space group, the equivalent atom positions of the octahedral Fe ions shown in Table S2.

$$
\begin{aligned}
\boldsymbol{F_{200}} = {} & f_1\cdot\exp[2\pi i\cdot 2\cdot 1/2] + f_1\cdot\exp[2\pi i\cdot 2\cdot 1/2] + f_1\cdot\exp[2\pi i\cdot 2\cdot 0] + f_1\cdot\exp[2\pi i\cdot 2\cdot 0] + \dots \\
& f_2\cdot\exp[2\pi i\cdot 2\cdot 1/4] + f_2\cdot\exp[2\pi i\cdot 2\cdot 1/4] + f_2\cdot\exp[2\pi i\cdot 2\cdot 3/4] + f_2\cdot\exp[2\pi i\cdot 2\cdot 3/4] + \dots \\
& f_3\cdot\exp[2\pi i\cdot 2\cdot 3/4] + f_3\cdot\exp[2\pi i\cdot 2\cdot 3/4] + f_3\cdot\exp[2\pi i\cdot 2\cdot 1/4] + f_3\cdot\exp[2\pi i\cdot 2\cdot 1/4] + \dots \\
& f_4\cdot\exp[2\pi i\cdot 2\cdot 0] + f_4\cdot\exp[2\pi i\cdot 2\cdot 0] + f_4\cdot\exp[2\pi i\cdot 2\cdot 1/2] + f_4\cdot\exp[2\pi i\cdot 2\cdot 1/2] + \dots \\
= {} & 4 f_1 - 4 f_2 - 4 f_3 + 4 f_4 = 4(f_1 - f_2 - f_3 + f_4)
\end{aligned}
$$

$$
\begin{aligned}
\boldsymbol{F_{020}} = {} & f_1\cdot\exp[2\pi i\cdot 2\cdot 1/2] + f_1\cdot\exp[2\pi i\cdot 2\cdot 0] + f_1\cdot\exp[2\pi i\cdot 2\cdot 1/2] + f_1\cdot\exp[2\pi i\cdot 2\cdot 0] + \dots \\
& f_2\cdot\exp[2\pi i\cdot 2\cdot 3/4] + f_2\cdot\exp[2\pi i\cdot 2\cdot 1/4] + f_2\cdot\exp[2\pi i\cdot 2\cdot 3/4] + f_2\cdot\exp[2\pi i\cdot 2\cdot 1/4] + \dots \\
& f_3\cdot\exp[2\pi i\cdot 2\cdot 0] + f_3\cdot\exp[2\pi i\cdot 2\cdot 1/2] + f_3\cdot\exp[2\pi i\cdot 2\cdot 0] + f_3\cdot\exp[2\pi i\cdot 2\cdot 1/2] + \dots \\
& f_4\cdot\exp[2\pi i\cdot 2\cdot 1/4] + f_4\cdot\exp[2\pi i\cdot 2\cdot 3/4] + f_4\cdot\exp[2\pi i\cdot 2\cdot 1/4] + f_4\cdot\exp[2\pi i\cdot 2\cdot 3/4] + \dots \\
= {} & 4 f_1 - 4 f_2 + 4 f_3 - 4 f_4 = 4(f_1 - f_2 + f_3 - f_4)
\end{aligned}
$$

The presence of the (±200) reflections indicate $f_1 + f_4 \neq f_2 + f_3$ and the presence of the (0±20) reflections indicate $f_1 + f_3 \neq f_2 + f_4$. To meet these conditions, one of the atom form factors should be different with the other three atom form factors, that is, the valence charge in one subgroup is different with other subgroups. For example, in subgroup 1 $Fe^{(2.5+\delta1)+}$ is $Fe^{2+}$, and subgroups 2, 3, 4, the $Fe^{(2.5+\delta2)+} = Fe^{2.76+}$ in order to keep the system neutral. Based on this charge ordering model, the originally forbidden {200} reflections can be observable due to the rotational symmetry breaking. There are 4 subgroups on the octahedral sites, each subgroup can be chosen to be different with the other three subgroups, therefore there are four twin variants in total. In each twin variant only one three-fold rotational symmetry along <111> direction is preserved, and the other three rotational symmetries are broken as shown in Fig. S5.

The dynamic electron diffraction simulation result is shown in Fig. S6. The intensity ratios of the reflections ($I_{200}$:$I_{240}$:$I_{600}$) in the experiment and simulation is summarized in Fig. 1g. The sample thickness for the simulation is determined to be 60 nm. The values of valence states on Fe ions are 2.54 and 2.486 for the simulation.

Structure factors of {200} reflections in atomic displacements models

With regards to atomic displacements model, the 16*d* Fe ions are separated in the same way as in the charge ordering model. That is, 16*d* site symmetry is broken, the face centered translational symmetry is preserved.

In the atomic displacements model shown in Table S3, we assume that there is no charge ordering on the Fe ions firstly. The atom form factor is *f*. Therefore, the structure factor of (200) reflection can be written:

$$\begin{aligned}\boldsymbol{F_{200}} = {} & f\cdot\exp[2\pi i\cdot 2\cdot(1/2+\delta_1)] + f\cdot\exp[2\pi i\cdot 2\cdot(1/2+\delta_1)] + f\cdot\exp[2\pi i\cdot 2\cdot\delta_1] + f\cdot\exp[2\pi i\cdot 2\cdot\delta_1] + \ldots \\ & f\cdot\exp[2\pi i\cdot 2\cdot(1/4+\delta_2)]+f\cdot\exp[2\pi i\cdot 2\cdot(1/4+\delta_2)]+ f\cdot\exp[2\pi i\cdot 2\cdot(3/4+\delta_2)] + f\cdot\exp[2\pi i\cdot 2\cdot(3/4+\delta_2)] + \ldots \\ & f\cdot\exp[2\pi i\cdot 2\cdot(3/4+\delta_2)]+f\cdot\exp[2\pi i\cdot 2\cdot(3/4+\delta_2)] + f\cdot\exp[2\pi i\cdot 2\cdot(1/4+\delta_2)] + f\cdot\exp[2\pi i\cdot 2\cdot(1/4+\delta_2)] + \ldots \\ & f\cdot\exp[2\pi i\cdot 2\cdot\delta_2] + f\cdot\exp[2\pi i\cdot 2\cdot\delta_2] + f\cdot\exp[2\pi i\cdot 2\cdot(1/2+\delta_2)] + f\cdot\exp[2\pi i\cdot 2\cdot(1/2+\delta_2)] \\ = {} & 4 f\cdot[\exp(2\pi i\cdot 2\cdot\delta_1) - \exp(2\pi i\cdot 2\cdot\delta_2)]\end{aligned}$$

To ensure $\boldsymbol{F_{200}}$ is different from zero, $\delta_1 \neq \delta_2$.

The dynamic electron diffraction simulation result is shown in Fig. S7. The atomic displacement is $0.005\boldsymbol{a_c}$ (~ 0.042 Å), which is larger than the spatial resolution of neutron scattering in(*7, 10*). However, the "forbidden" peaks, e.g., {200} and {600}, are not observed in the neutron scattering experiments, which indicate the extra "forbidden" peaks are mainly from electronic structure, i.e., charge ordering phase.

Space group and point group analysis based on the CO model

The point group of $Fd\bar{3}m$ space group is $O_h$, which has 48 symmetry operations. The charge ordering structure by splitting atomic coordinates on *16d* site reduces the symmetry operations from 48 to 12. According to the maximum non-isomorphic subgroups in $Fd\bar{3}m$ space group, the point group is reduced from $O_h$ to $D_{3d}$, and the corresponding space group is transformed from $Fd\bar{3}m\ (No. 227)$ to $R\bar{3}m\ (No. 166)$. As we mentioned above, there are four types of charge distribution model. In each type of model, one three-fold rotational symmetry along <111> directions is preserved, the rotational symmetry in the other three directions are broken.

CO peak intensity and diffuse scattering intensity comparison

In the *in situ* cooling TEM experiment, we observed diffuse scattering signal right above phase transition temperature $T_v$. The diffraction pattern shown below in Fig. S8 is taken at 121 K from Sample 1. The sharp "forbidden" peaks from the monoclinic phase become smear and turn to diffuse. Simultaneously, the charge ordering peaks are still observable. The intensities of the diffuse scattering signal and charge ordering peaks are comparable. However, the neutron study in(*7, 10, 11*) detected the diffuse scattering signal but did not observe charge ordering peaks, which indicates electronic ordering contributes more to the {200} reflections in the high-temperature cubic phase.

Form factor calculation for electron scattering and X-ray scattering

In order to compare the sensitivity to the valence electron on Fe ions, we calculated to the form factor $Fe^{2+}$ and $Fe^{3+}$ form factors as a function of scattering vector, the calculated form factor squared ($f^2$) are shown in Fig. S9. At the scattering vector of {200} reflection, $\boldsymbol{q}$ = 0.119 Å$^{-1}$, the $f^2$ difference between $f_e^2(Fe^{2+})$ and $f_e^2(Fe^{3+})$ is much larger than that between $f_x^2(Fe^{2+})$ and $f_x^2(Fe^{3+})$, which illustrates electron scattering is more sensitive to the valence charge disproportion at small scattering angles.

CO peak intensity as a function of valence charge on Fe ions

Figure 2b shows that the intensity of CO peak increases upon cooling and reaches a maximum value around the phase transition temperature. To figure out the intensity variation in the cubic phase, which is mainly related to the charge ordering, we calculated the intensity variations by changing the charge discrepancy on the Fe ions. The calculated result is shown in Fig. S10. When the charge is uniformly distributed on the Fe ions ($Fe_{oct.}^{2.5+}$), that is, no charge ordering exists ($\delta_1$= $\delta_2$ = 0), the CO peak intensity is zero. The CO peak shows up, when the charge discrepancy exists among the Fe ions shown in Fig. S10b. The calculated result also demonstrates that the larger charge discrepancy, the higher intensity on CO peak.

*In situ* heating TEM experiment

To explore whether the nematic phase can be thermally removed at high temperature, Sample 1 was gradually heated up from room temperature to 851 K and the electron diffraction patterns were captured at 325 K, 349 K, 372 K, 423 K, 625 K, 758 K and 851 K, respectively. The diffraction patterns at a few temperatures are shown in Figs. S11a-S11d. The CO peaks can be observed until 423 K and there are some extra peaks start to show up at 625 K. Additionally, the diffuse scattering signal is enhanced in the background at higher temperatures. we infer that the sample could be damaged by the heating effect, due to the sample being thin susceptible to stoichiometry changes at high temperatures. We analyzed the electron diffraction patterns taken below 625 K and the results are shown in Fig. S11e. Upon warming, the CO peak intensity gets reduced and the peak width gets broader, demonstrating the charge disproportion between Fe ions is smaller and the coherent length is decreased due to the thermal effect. Based on the *in situ* heating experiment, we deduce that the hypothetical temperature at which the nematicity disappears is higher than 423 K and does not necessarily exist at all.

Two possible scenarios in the Verwey transition upon cooling

The precursor behaviors in the Verwey transition have been observed and discussed, which presented as diffuse scattering in the diffraction techniques, such as, electron diffraction, synchrotron x-ray diffraction and neutron scattering (*10–12*, *62*). The temperature range of the diffuse scattering observation highly depends on the sample quality and the stoichiometry. It is worthy of mentioning that we observed these diffuse scattering as well on the electron diffraction patterns from our well calibrated $Fe_3O_4$ samples: above the Verwey temperature, the diffuse scattering signals have significantly broadened and weak intensity distribution at barely defined $\boldsymbol{q}$ position similar (but not identical) to those reflections from the low-temperature trimeron phase. Hitherto, the origin of the diffuse scattering is still a highly debated issue that could arise from multiple possibilities, such as a short-range trimeron order. A x-ray pair distribution function study demonstrated that the structural distortion within Third Unit Cell range was found to immediately disappear when the temperature is increased across the Verwey transition, while the local structural distortion within First Unit Cell range appears to exist to high temperature of about 800 K (*13*). However, it is still challenging for us to accurately interpret the nature of the structural distortions and the short-range trimeron order.

The discovery of the electronic nematicity at the temperatures above the Verwey transition is the center of the manuscript. While the precursors of trimerons may coexist with nematic reflection over a range of temperatures, the evolution from nematicity to trimerons involves a more fundamental change in the underlying electronic structure of the material. As shown in Fig. S12, two phase-transition scenarios upon cooling are compared. Whether electronic nematic phase exists at the temperatures above the Verwey transition would result in distinctive experimental observations across the Verwey transition with fundamentally different driving mechanisms from theoretical perspective. For example, electronic nematicity (rhombohedral; cubic symmetry is already broken) can help stabilize trimeron order (monoclinic) by providing a preferred directionality for the electronic degrees of freedom that participate in trimeron formation.

On the experimental side, if we perform thermal cycling in the vicinity of the Verwey transition, that preferred direction locally in the monoclinic phase is randomly selected every time when the system is cooled down below the Verwey temperature without long-range nematicity, while the preferred direction is pre-selected by the nematic order locally through each thermal cycling. Of course, in a realistic experiment, other effects such as strain and defects could also help to pre-select the preferred direction in the monoclinic phase. Nevertheless, for those two scenarios in Fig. S12, the Verwey transition involves very different thermal fluctuations and thus they must belong to two distinct families of phase transitions, as will be shown below from the theory side.

From the theoretical point of view, the two scenarios in Fig. S12 belong to two different universality classes and they are described by different field theories. This can be demonstrated by an analogy. We can think of the trimeron ordered insulating state as a solid (crystal or smectic) formed by electrons, while the high-temperature cubic state is an electron liquid (*34*). In the top scenario in Fig. S12, the Verwey transition corresponds to a direct solid-liquid transition. In contrast, in the bottom scenario in Fig. S12, the Verwey transition corresponds to a transition between a solid/smectic and nematic fluid. From the study of classical liquid crystals, the bottom scenario is known to be highly unconventional, where nematic order parameter serves the role of a gauge field (*37*, *63*, *64*).

In summary, our finding of the electronic nematicity reveals that the Verwey transition belongs to a new universality class, different from the conventional picture. The substantial impact of the electronic nematic phase can be demonstrated experimentally and theoretically for understanding the driving mechanisms of the Verwey transition.

Charge order arrangement in cubic phase and monoclinic phase

In the cubic phase, there are four atomic layers of octahedral Fe ions in one-unit cell along the ***c*** axis in the charge ordering models. Below $T_v$, the unit cell is expanded into $\sqrt{2a} \times \sqrt{2a} \times 2a$ in the monoclinic phase with trimeron order. Thus, there are eight atomic layers of octahedral Fe ions along the ***c*** axis in the monoclinic phase. To simplify the charge distribution sketch, the valence states of Fe ions are rounded to the nearest integer in the low-temperature phase, as shown in (*2*). The charge distribution in each layer in the electronic nematicity phase and trimerons phase is shown in Figs. S13 and S14, respectively.

Intensity variation Bragg peaks during photoexcitation (@30K, pump fluence 4 $mJ\cdot cm^{-2}$)

The MeV UED experiment was taken at temperature ~ 30 K, where the structure is monoclinic with the trimeron charge ordering. We prepared a UED single-crystal sample at the [110] zone axis for intensity measurements from more Bragg / SL peaks than the [001] zone shown in Fig. 1 and Fig. 2. During the UED experiment, in order to enhance the SLs intensity, the sample was slightly tilted off the zone axis on purpose as shown in Fig. S15. All the SL peak intensities drop quickly in the first 700 fs while no obvious change can be measured from Bragg peak intensities. A slow dynamic process follows, which starts from ~ 700 fs and lasts at least a few ps in our observations, which is consistent with the timescale observed in the ultrafast x-ray diffraction study (*47*).

Based on the signal noise ratio of intensity change on each Bragg peak, we found that the intensity variations can be classified as two typical types: one type is intensity decreasing (marked in blue color); another type is intensity increasing (marked in yellow color),which are labeled in Fig. S15. The measured peaks are numbered. The equivalent $(hkl)$ and $(\bar{h}\bar{k}\bar{l})$ reflections are labeled using the same number.

The temporal evolutions of the Bragg peaks are shown below in Fig. S16. These Bragg peaks are selected considering the signal noise ratio.

Dynamic electron diffraction simulation for UED results

To understand the peak intensity variations, we performed dynamic electron diffraction analysis considering atomic charge distribution, atomic displacements, and existence of twin variants. Since the UED sample was cut from a bulk material and thinned down to less than 100 nm using Focused Ion Beam, the thickness of the sample is larger than that of the typical cleaved 2D materials and the bulk sample could be relatively flat. That is, the precession effect is reduced, and the multiple scattering effect is enhanced. Therefore, we performed dynamic electron diffraction simulation using Bloch wave method. Due to the multiple twin variants captured in the UED experiment, we considered the Bragg peaks coming from each twin structure. Since the diameter of electron probe is around 100 μm, we assume the fraction of domains is fixed and evenly distributed in the electron diffraction calculation. All the peaks considered in the simulation are listed in Table S4 and Table S5. Each reflection in the UED pattern is numbered, e.g., B1 (Bragg peak 1), SL1 (superlattice peak 1).

A. Intensities of Bragg peaks and SLs with different valence state of Fe ions on octahedral site

The charge ordering model for the electron diffraction calculation is based on the trimeron model at low temperature. We tuned the charge discrepancy of Fe ions in the trimeron, e.g., 2.6+, 2.4+; 2.7+, 2.3+, etc., and calculated the corresponding intensities of all the measured Bragg peaks and SLs. The calculated results are shown in the plot in Fig. S17. The calculated result indicates that charge discrepancy has notable impact on the intensity of SLs, compared with the intensity variation of Bragg peaks. The diffraction calculations reveal that the intensity change of the SL peaks mainly comes from the change in electronic order parameter and the intensity variations of the Bragg peaks can be mostly attributed to atomic displacement, i.e., lattice distortions.

B. Intensities of Bragg peaks induced by phonon modes

According to the literature research, Refs(*40*, *41*, *47*) pointed that there are three dominant phonon modes in the cubic phase involved in the phase transition: $X_3$ (transverse optical phonon); $\Delta_5$ (transverse acoustic phonon) and $\Gamma_5$. $X_3$ and $\Delta_5$ phonon modes correspond to the primary order parameters for the structural phase transition. To investigate each specific phonon effect in the monoclinic phase, we built atomic displacement model according to each frozen phonon mode and calculated intensities of Bragg peaks and SLs shown in Figs. S18-S20. Since these phonon modes with $X_3$ and $\Delta_5$ symmetry remarkably distort the octahedral sites, in atomic distortion models, we only focus on the Fe ions located in the octahedral sites, i.e., involved in the trimeron ordering, so the Fe ions on the tetrahedral sites are omitted.

Calculation result based on $\Gamma_5$ phonon mode Figure S18 is the intensity variation induced by the $\Gamma_5$ phonon mode, which is related to the β angle changes from 90º to 90.23º without affecting the relative atomic positions in the unit cell during the phase transition as shown in Fig. S18a. In the calculation, the β angle in the monoclinic phase is changed from 90.23º to 90º step by step. The calculation result indicates that β angle variation has little impact on the reflection intensities. Figure S18b is a representative result calculated for Bragg peak 1 and SL 1. Hence $\Gamma_5$ phonon mode is not the dominant mode during the photoexcitation process.

Calculation result based on $\Delta_5$ phonon mode Figure S19a shows atomic displacements corresponding to $\Delta_5$ phonon mode. For the Fe ions on the octahedral site, there are total 8 Fe-O layers in one unit cell in the monoclinic phase. The right part in Fig. S19a shows the relative amplitude of the atomic displacement in each layer. We calculated the intensities of Bragg peaks after moving the atoms off the original positions. The calculated intensity variation due to atomic displacement is in Figs. S19b and S19c. The calculated intensity for each Bragg peak based on the original monoclinic phase is the reference for the intensity normalization. The Bragg peaks present different responses to the atomic displacements in the calculation. The intensities of the Bragg peaks shown in Fig. S19b are increased after photoexcitation in our experimental measurement, but the calculated intensities of Bragg peaks #5 and #16 decrease. Likewise, the intensities of the Bragg peaks shown in Fig. S19c decrease in the experimental observation. Calculated intensity of Bragg peak #30 increases, which is not consistent with the experiment. Therefore, we can conclude that $\Delta_5$ phonon mode is not the major contribution to the intensity changes after photoexcitation.

Calculation result based on $X_3$ phonon mode The calculated intensities for the Bragg peaks based on the $X_3$ phonon mode are shown in Fig. S20. All the trends of intensity variations are consistent with the measurement results. Hence, we can conclude that $X_3$ phonon mode is highly excited through the strong electron-phonon coupling upon photoexcitation.

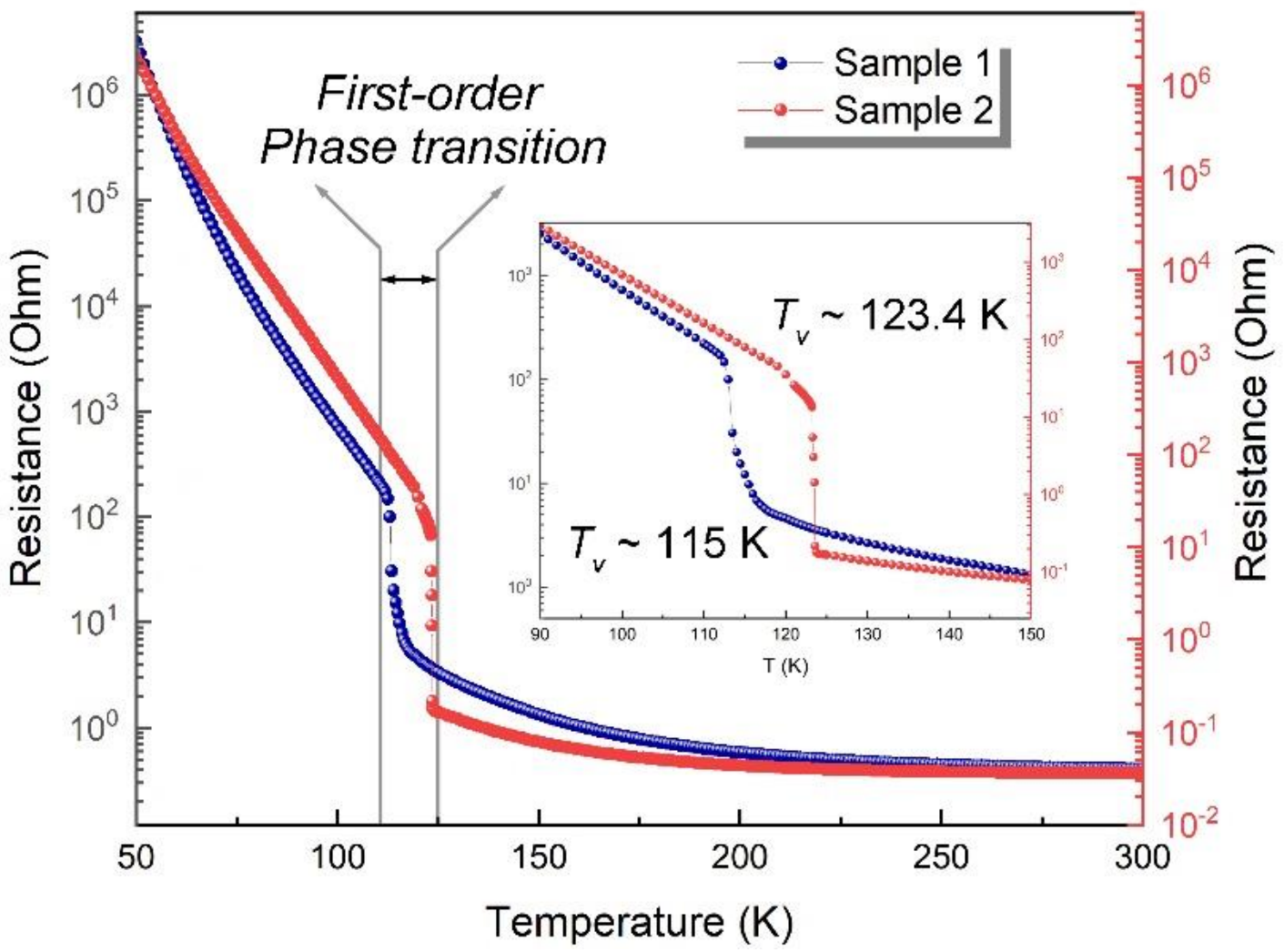

**Fig. S1.**

**Four probe resistance measurements for Sample 1(commercial sample, SurfaceNet GmbH, Germany) and Sample 2 (from Komarek group).** The samples were measured at zero field. The inset is an enlarge view of data in the vicinity of phase transition. $T_v$ in Sample 1 is ~ 115 K, and $T_v$ in Sample 2 is ~ 123.4 K. The measurement results demonstrate that the phase transitions in both samples are first-order phase transition, i.e., Verwey transition.

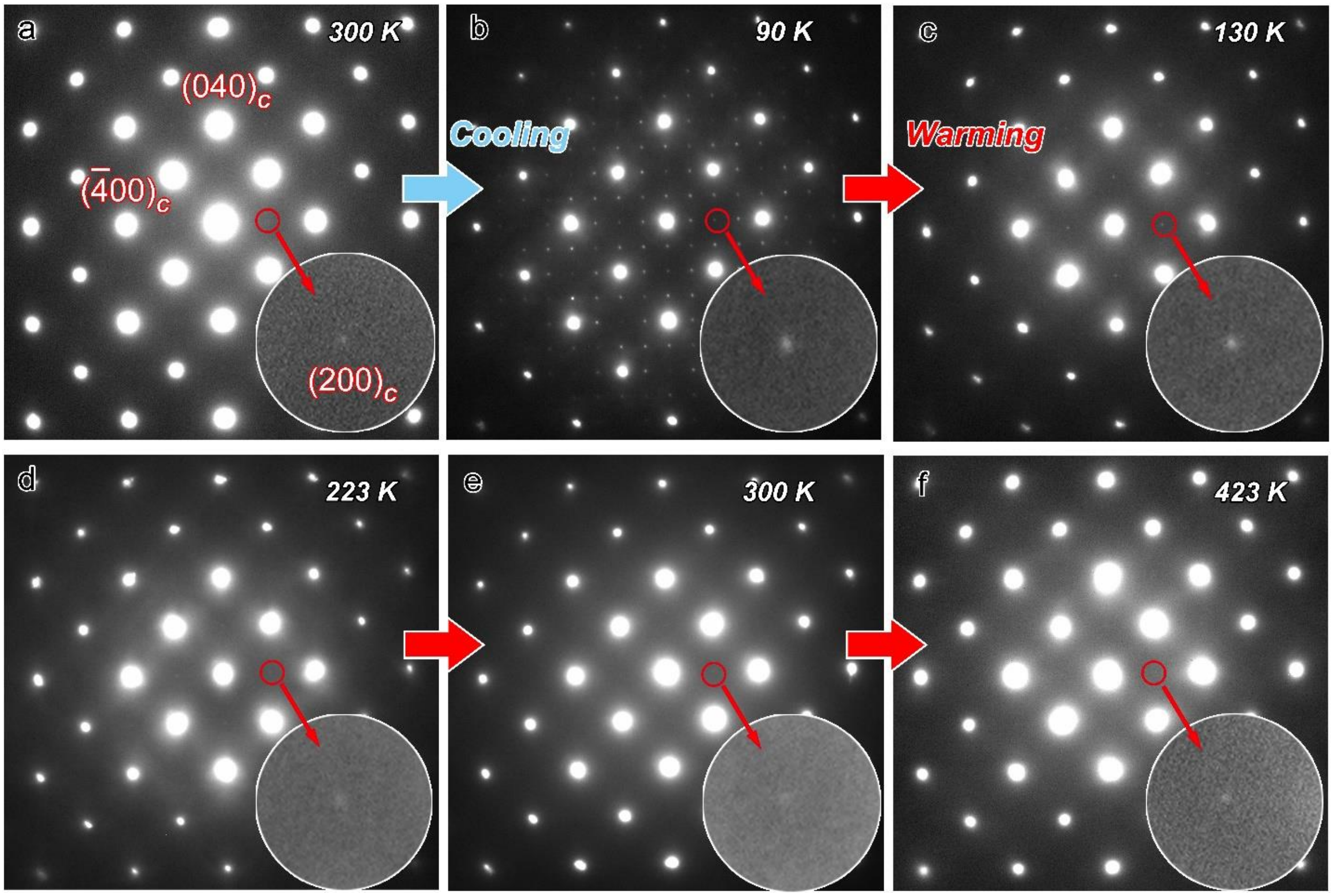

**Fig. S2.**

**A series of electron diffraction patterns of Sample 1 along <001> zone axis taken @ 90 K, 130 K, 223 K, 300 K and 423 K. a** Electron diffraction pattern taken at room temperature before cooling down, from the inverse spinel cubic phase; **b** Electron diffraction pattern from the monoclinic phase; **c-f** Diffraction patterns captured during the warming process. The inserts are the enlarged (200) peak, which is survived above the phase transition temperature.

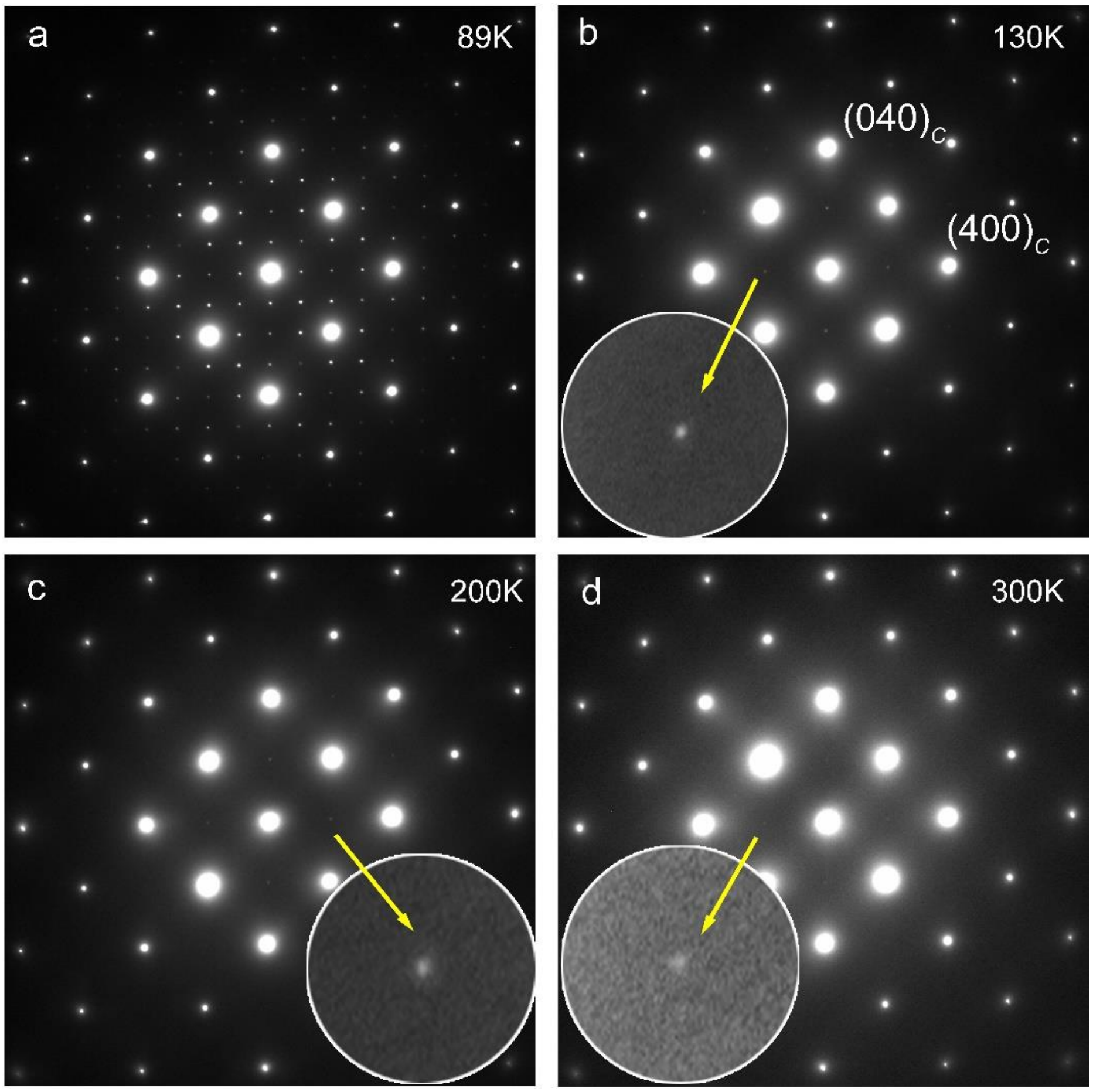

**Fig. S3.**

***In situ* cooling TEM experiment for Sample 2. a-d** Electron diffraction patterns along <001> zone axis during the warming process. The inserts are the persistent {200} peaks above the phase transition.

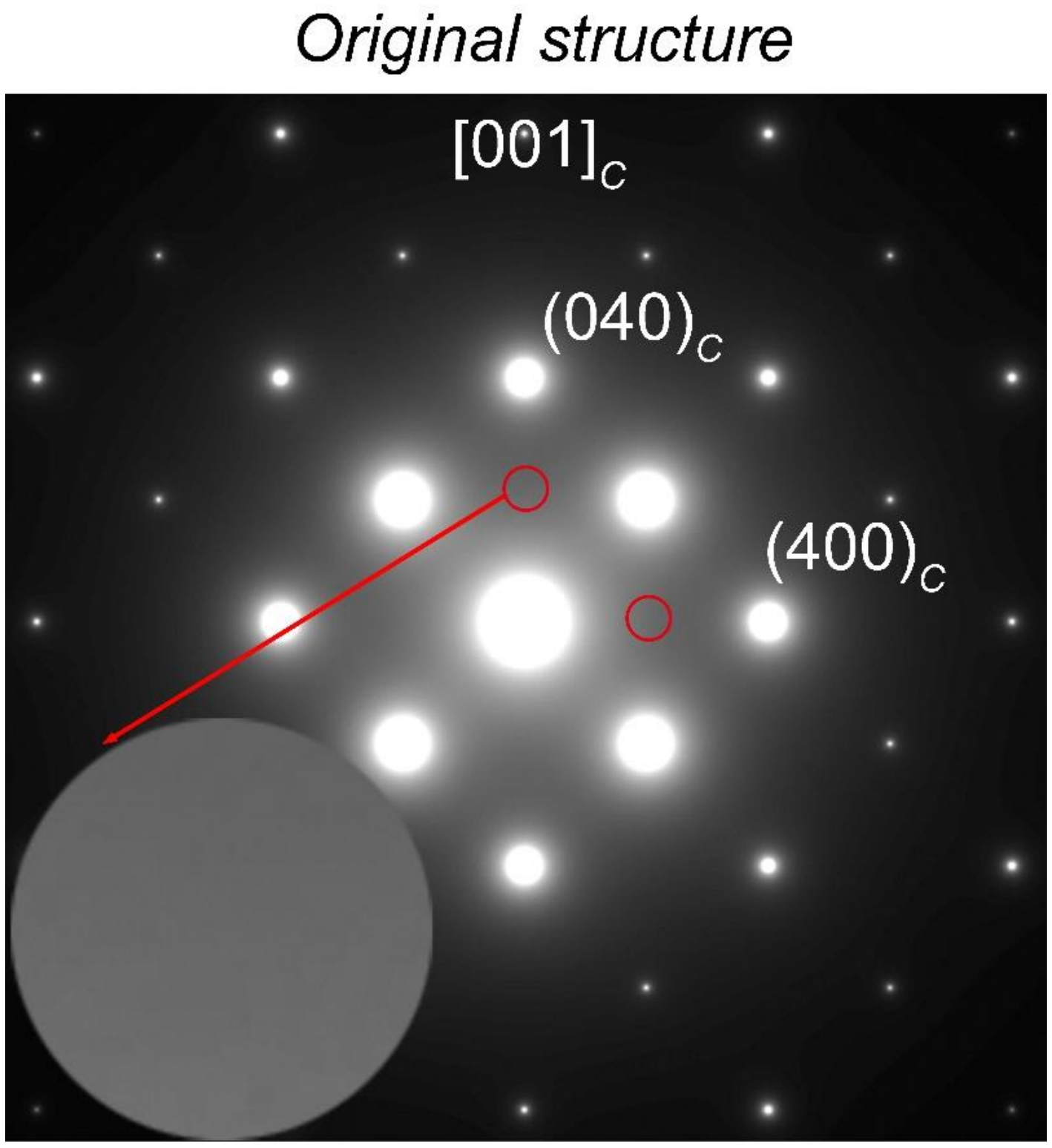

**Fig. S4.**
**Dynamic electron diffraction simulation including the appropriate sample thickness and multiple scattering effect.** Electron diffraction calculation from the standard cubic spinel structure in $Fe_3O_4$ along [001] direction. {200} reflections are still forbidden.

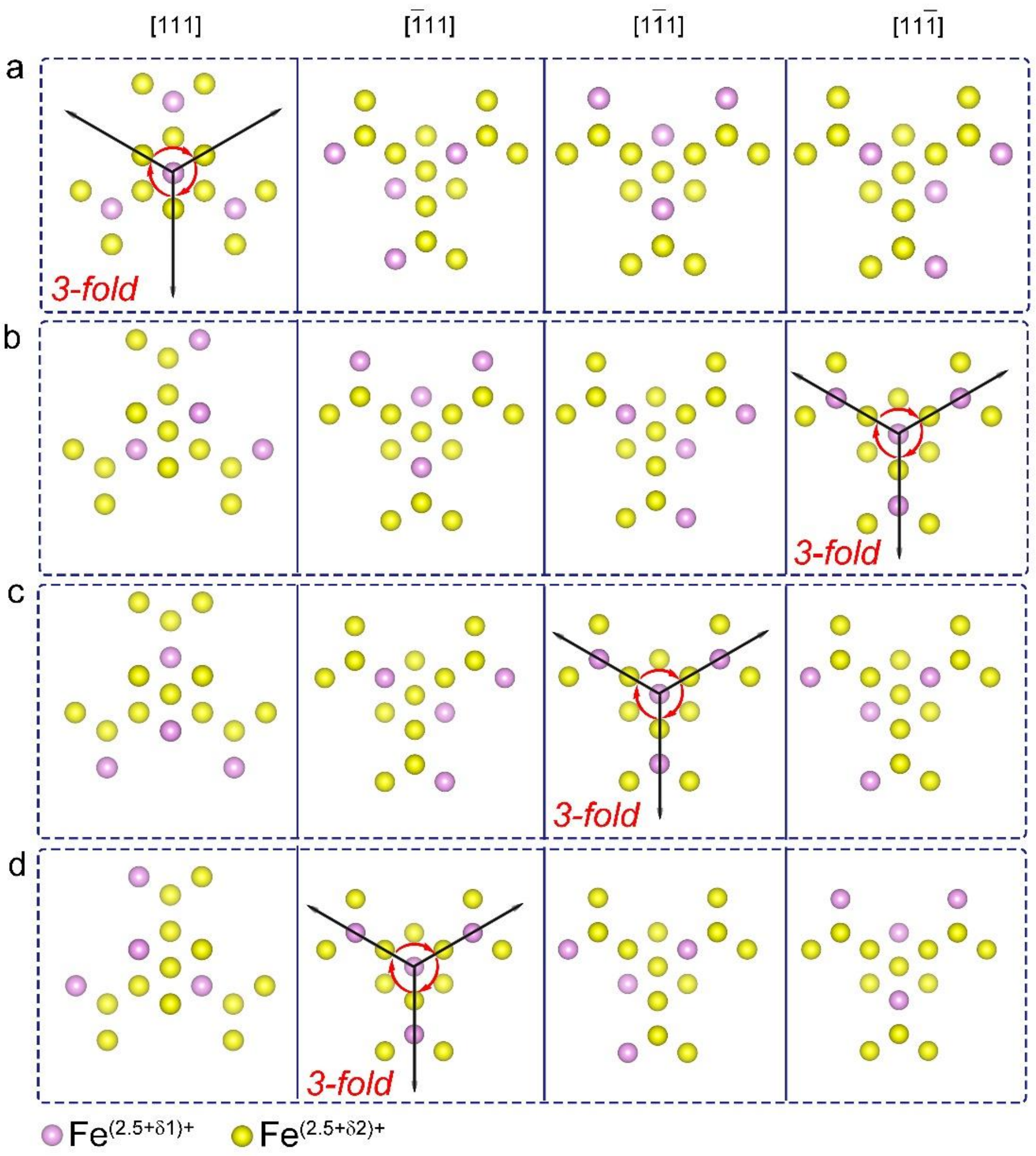

**Fig. S5.**

**Charge distribution along <111> directions in four variants. a** The valence state with $Fe^{(2.5+\delta1)+}$ is on subgroup 1 (1/2, 1/2, 1/2), the valence state of Fe ions on subgroup 2 (1/4, 3/4, 0), subgroup 3 (3/4, 0, 1/4), and subgroup 4 (0, 1/4, 3/4) is $Fe^{(2.5+\delta2)+}$. **b-d** The valence state with $Fe^{(2.5+\delta1)+}$ is on (1/4, 3/4, 0), (3/4, 0, 1/4), (0, 1/4, 3/4) subgroups, respectively. The valence state with $Fe^{(2.5+\delta2)+}$ is on the other three subgroups. In each charge ordering model, only one three-fold rotational symmetry is preserved along <111> directions, which is highlighted. Only Fe ions on octahedral sites are shown in the models.

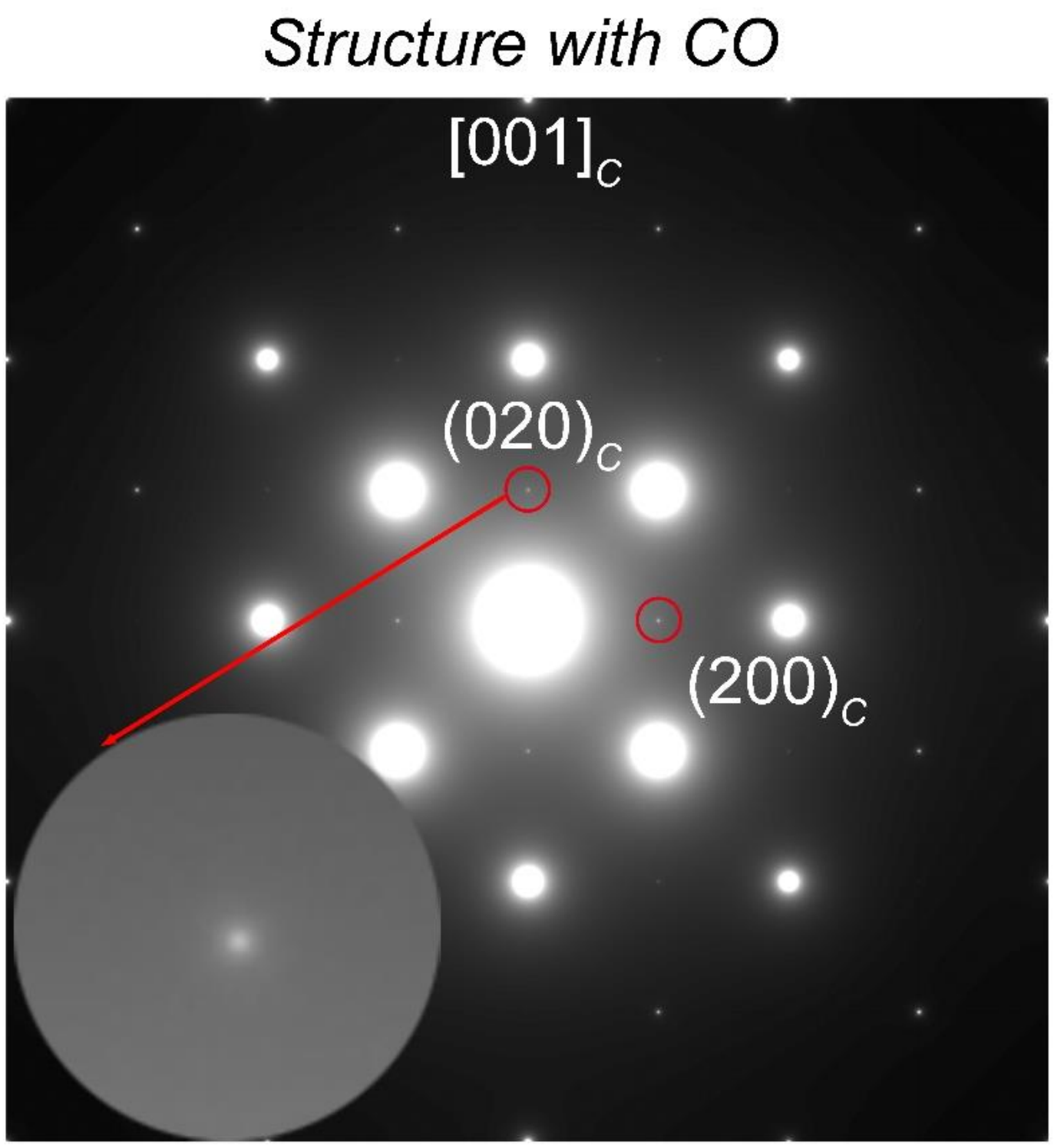

**Fig. S6.**
**Electron diffraction simulation result.** Dynamic calculation result based on the charge ordering model, (200) and (020) reflections are marked by red circles.

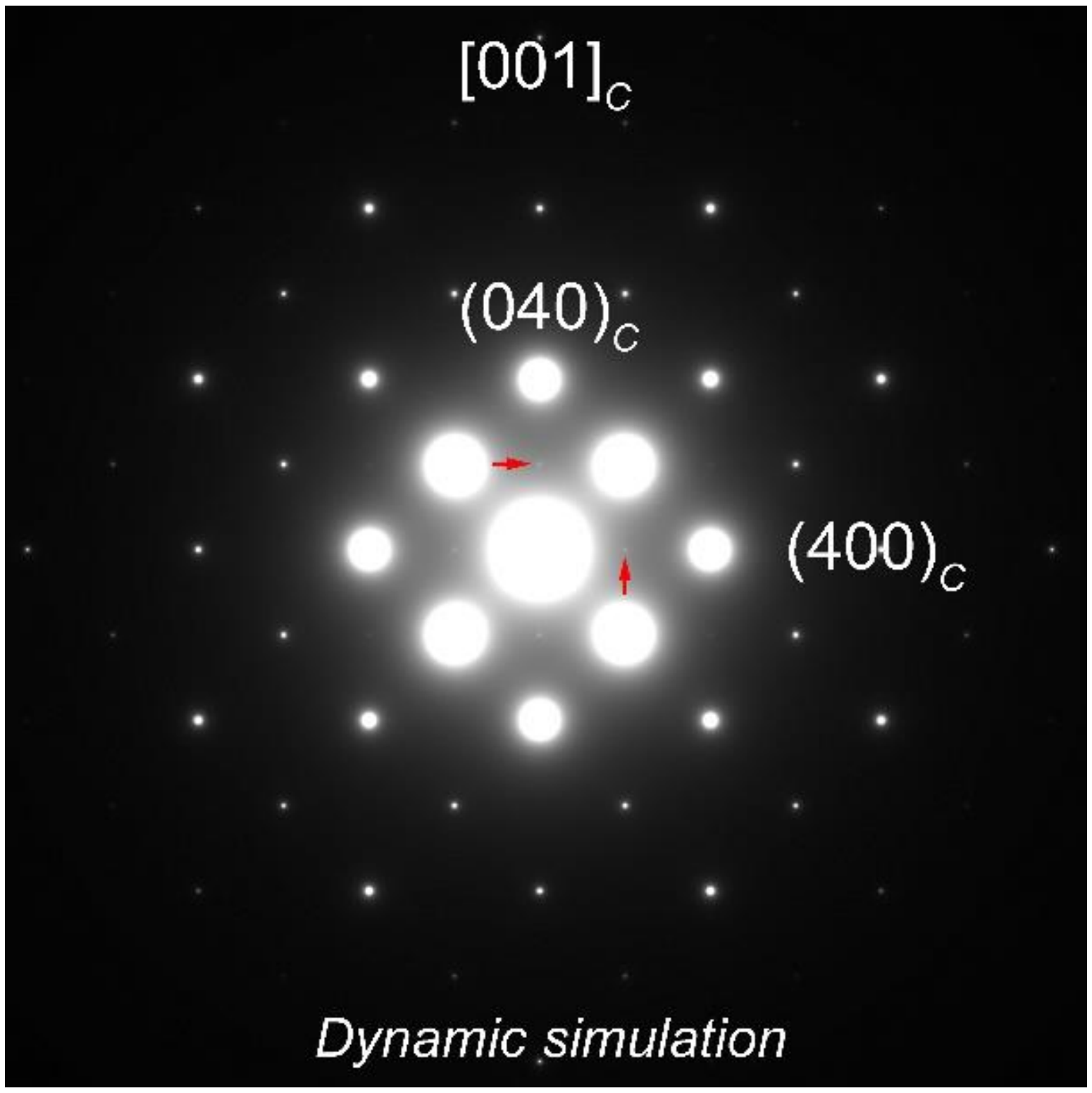

**Fig. S7.**

**Electron diffraction simulation of atomic displacement model.** The displacement is 0.005$\boldsymbol{a_c}$ and the sample thickness is 60 nm for the dynamic simulation.

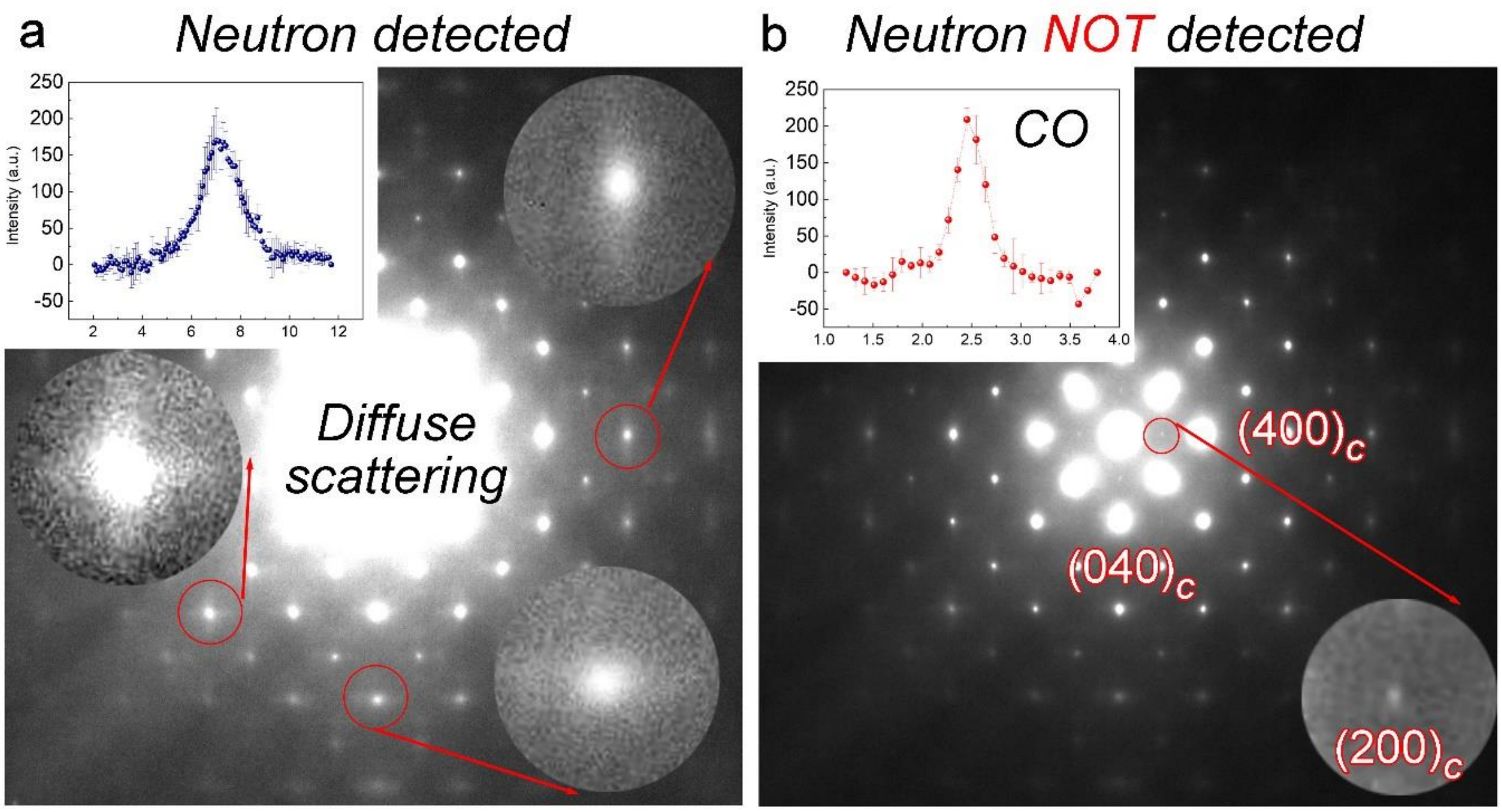

**Fig. S8.**

**Electron diffraction pattern taken slightly above $T_v$. a** and **b** patterns are the same diffraction pattern by changing image contrast. The diffuse scattering signal is observable in **a**. Diffuse scattering signals with different shapes are enlarged in the inserts. The insert plot is the averaged line profile intensity from these diffuse scattering signals. **b** shows the charge ordering {200} reflections. The averaged line profile intensity of these four peaks is inserted. The diffraction patterns were taken from Sample 1.

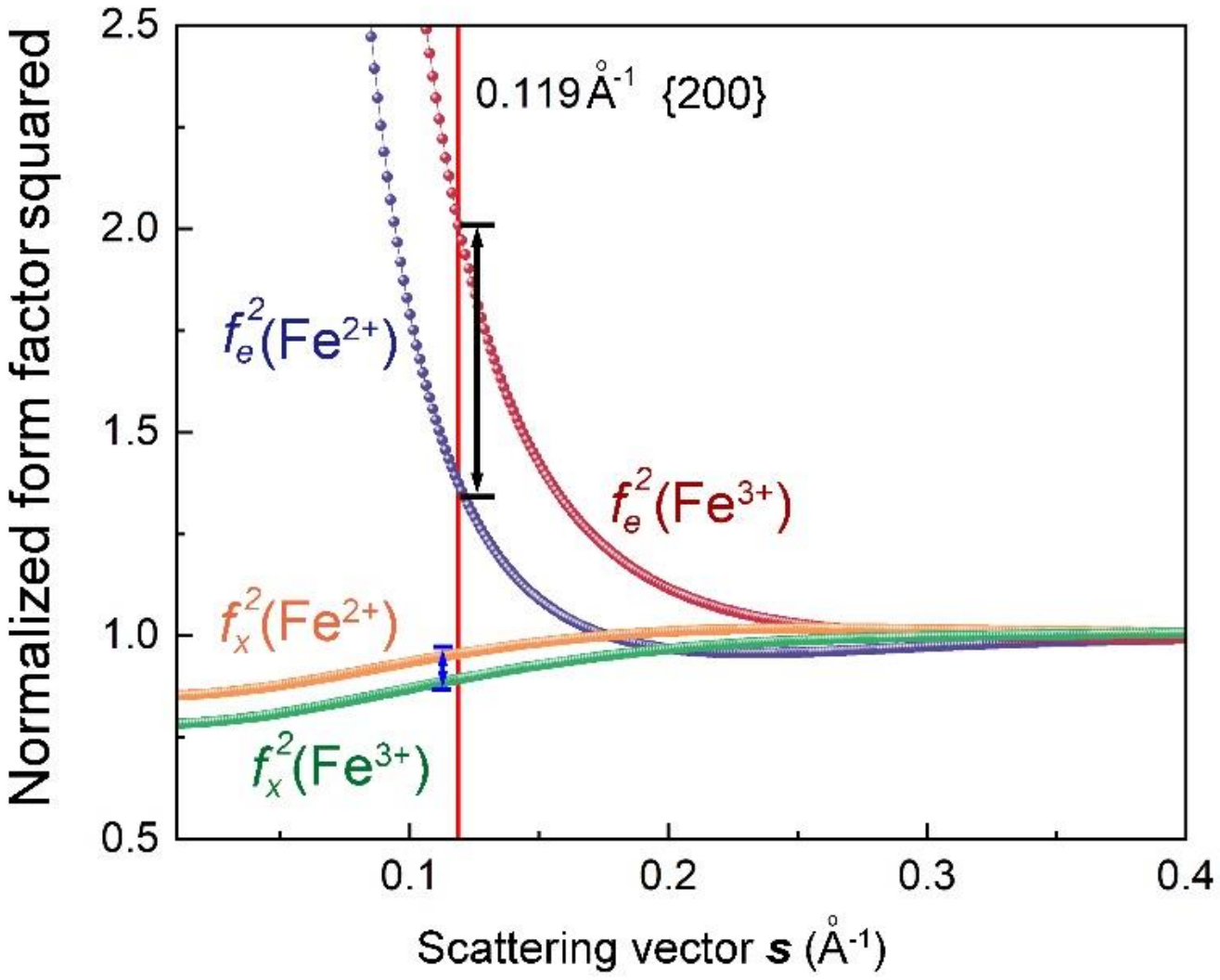

**Fig. S9.**

**Calculated electron form factor $f_e$ and x-ray form factor $f_x$ for $Fe^{3+}$ and $Fe^{2+}$.** The normalized form factor squared (normalized by the form factor of neutral Fe atom ($Fe^0$) are plotted as a function of scattering vector, ***s*** (Å$^{-1}$). The vertical red line indicates the scattering vector of {200}. The difference of form factor squared between $f_e^2(Fe^{2+})$ and $f_e^2(Fe^{3+})$ is marked by the black arrows, and that of $f_x^2(Fe^{2+})$ and $f_x^2(Fe^{3+})$ by the blue arrows. This difference illustrates the high sensitivity of electrons to valence charge at small scattering vectors.

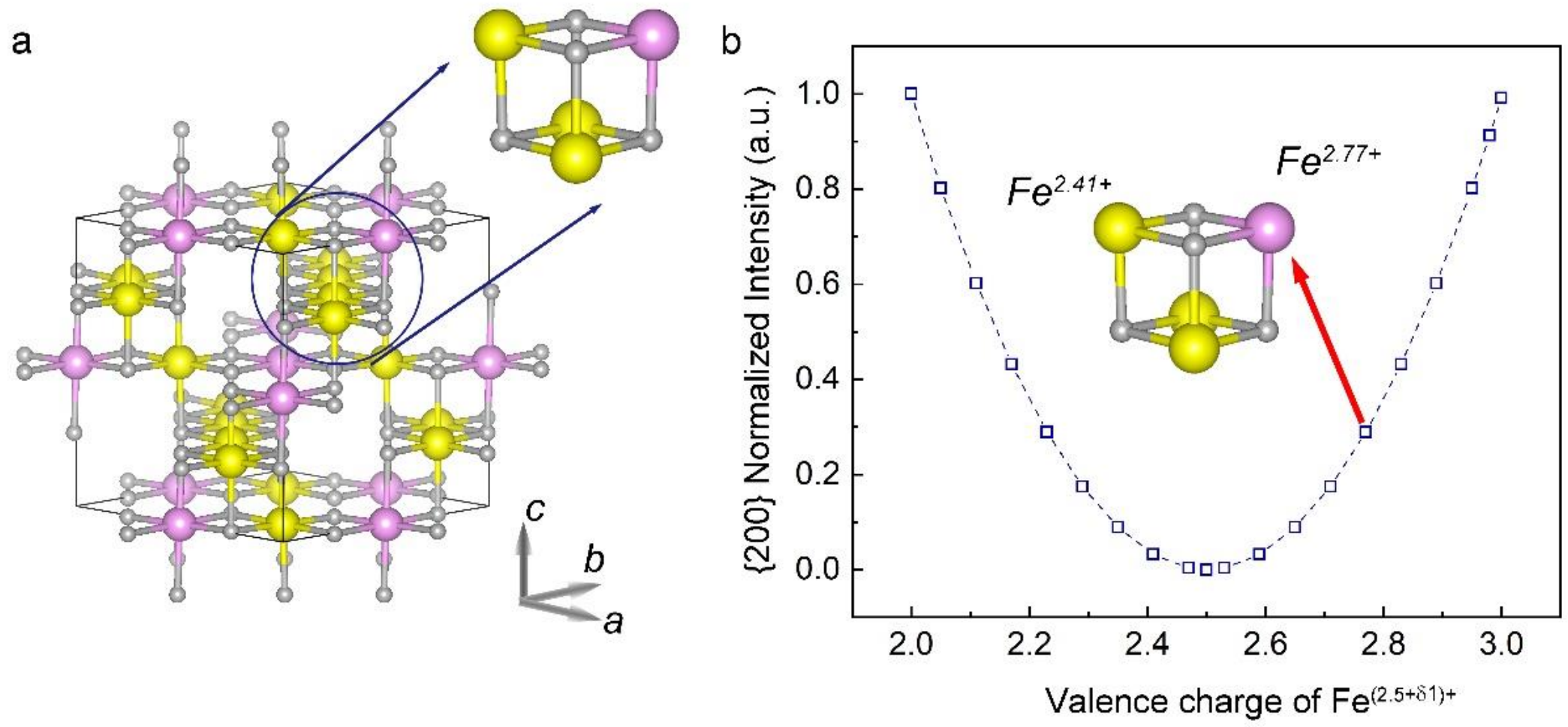

**Fig. S10.**

**CO peak intensity calculation as a function of valence charge in cubic phase. a** Crystal structure in the cubic phase, pink and yellow spheres are the Fe ions on the octahedral sites with different valence states, the grey sphere is the oxygen. The Fe ions on the tetrahedral sites are omitted. **b** Calculation result: CO {200} peak intensity as a function of the valence charge $Fe^{(2.5+\delta 1)+}$ on Fe ions.

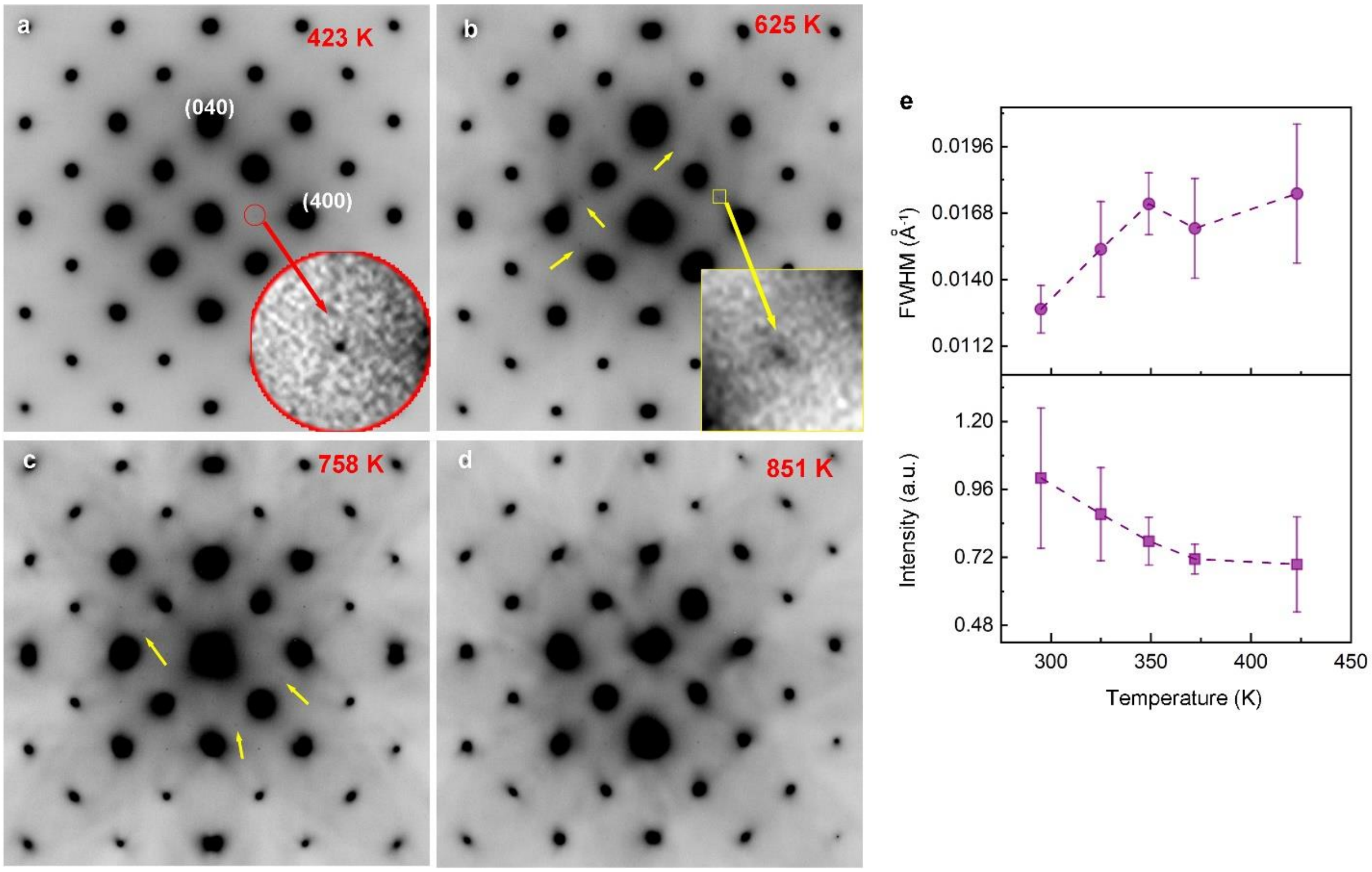

**Fig. S11.**

**Electron diffraction patterns in the *in situ* TEM heating experiment at a 423 K, b 625 K, c 758 K and d 851 K.** (200) reflection is shown in the insert in **a**. The insert in **b** is one of the extra peaks appearing at high temperatures. These extra peaks should not be observed in the cubic structure and appear upon warming. The arrows indicate the positions of the extra peak, which locates in the middle of two Bragg peaks. The arrows in **c** are used to highlight the extra peaks at 758 K. The diffuse scattering signal is strong in **c** and **d**. **e** Peak width and peak intensity of CO reflections as a function of temperature upon heating from 300 K, measured from Sample 1.

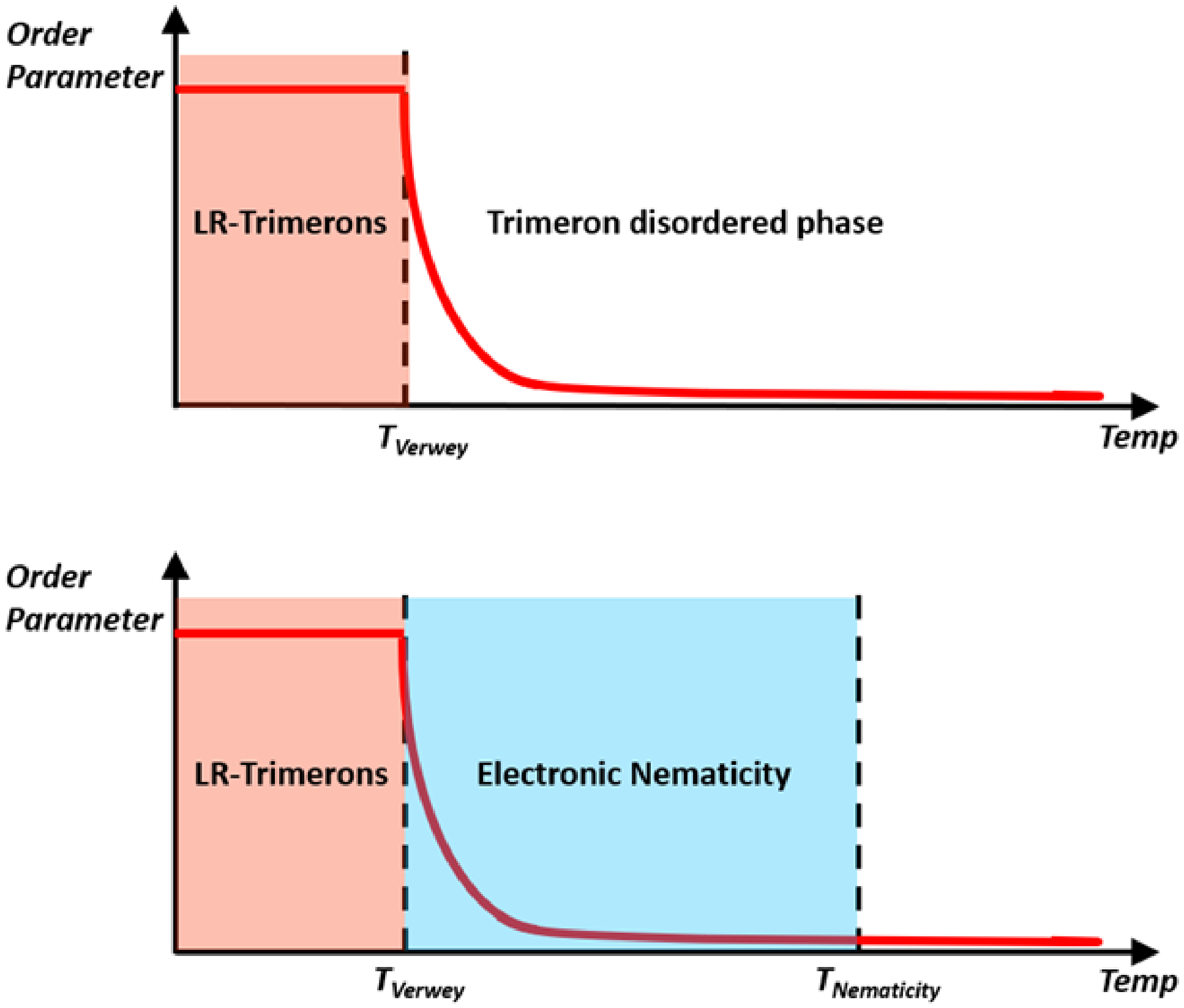

**Fig. S12.**

**A comparison of two possible phase-transition scenarios in $Fe_3O_4$ upon cooling.** Short-range trimerons may exist in a long-range trimeron-disordered phase with enhanced order parameter from short-range trimerons as temperature decreased (Top). Or short-range trimerons may exist in a long-range electronic nematic phase and both phases evolve with enhanced order parameters upon cooling (Bottom). Note that the plots of the order parameters (red lines) are only eye guides and may not reflect the real case quantitatively. These two scenarios would result in distinctive experimental observations across the Verwey transition with fundamentally different driving mechanisms from theoretical perspective.

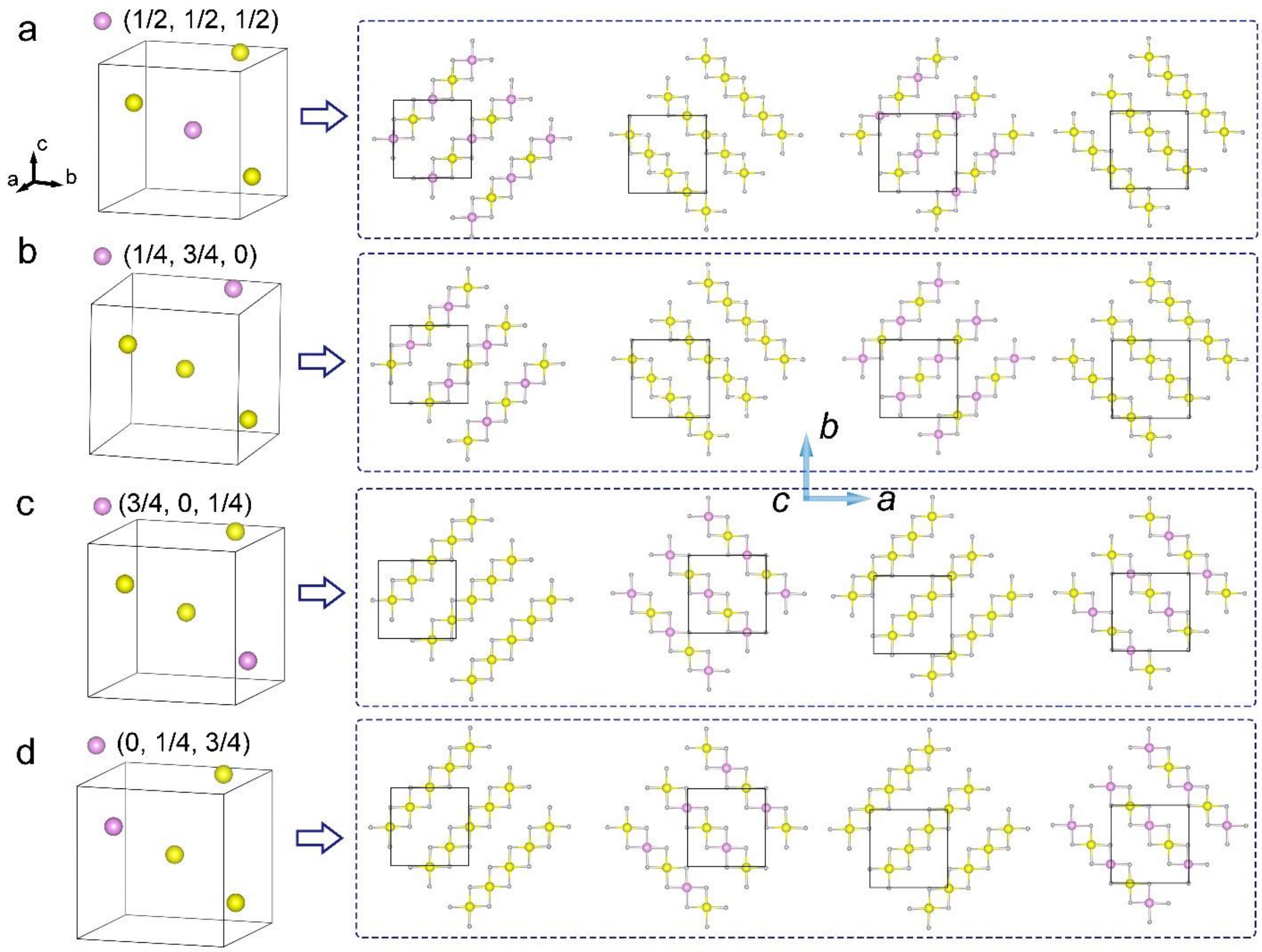

**Fig. S13.**

**Charge order arrangement in each layer in cubic phase. a-d** are four types of charge distribution. The left panel is one unit cell of cubic structure. The four Fe atoms are from octahedral sites. Other Fe atoms and oxygen atoms are omitted. The coordinates are (1/2, 1/2, 1/2), (1/4, 3/4, 0), (3/4, 0, 1/4), (0, 1/4, 3/4), which represent the four subgroups, mentioned in Table S2. In the charge ordering model, the valence charge in one subgroup is different with other three subgroups. **a** shows the valence charge of Fe atom at (1/2, 1/2, 1/2) is different with other three atoms, which is highlighted by pink color, and the valence charge is $Fe^{(2.5+\delta 1)+}$. The valence charge in the other three atoms is the same, i.e., $Fe^{(2.5+\delta 2)+}$, which is in yellow color. According to the face center symmetry in the cubic structure, each atom has another three equivalent atoms as shown in Table S2. There are total 16 Fe atoms in the octahedral sites, distributed in four layers in one unit cell. The unit cell is outlined by the black frame. The charge distribution in each layer along ***c*** axis is shown in the right panel. In **b**, **c** and **d**, the atom with the valence charge of $Fe^{(2.5+\delta 1)+}$ locates in 1/4, 3/4, 0), (3/4, 0, 1/4), (0, 1/4, 3/4), respectively.

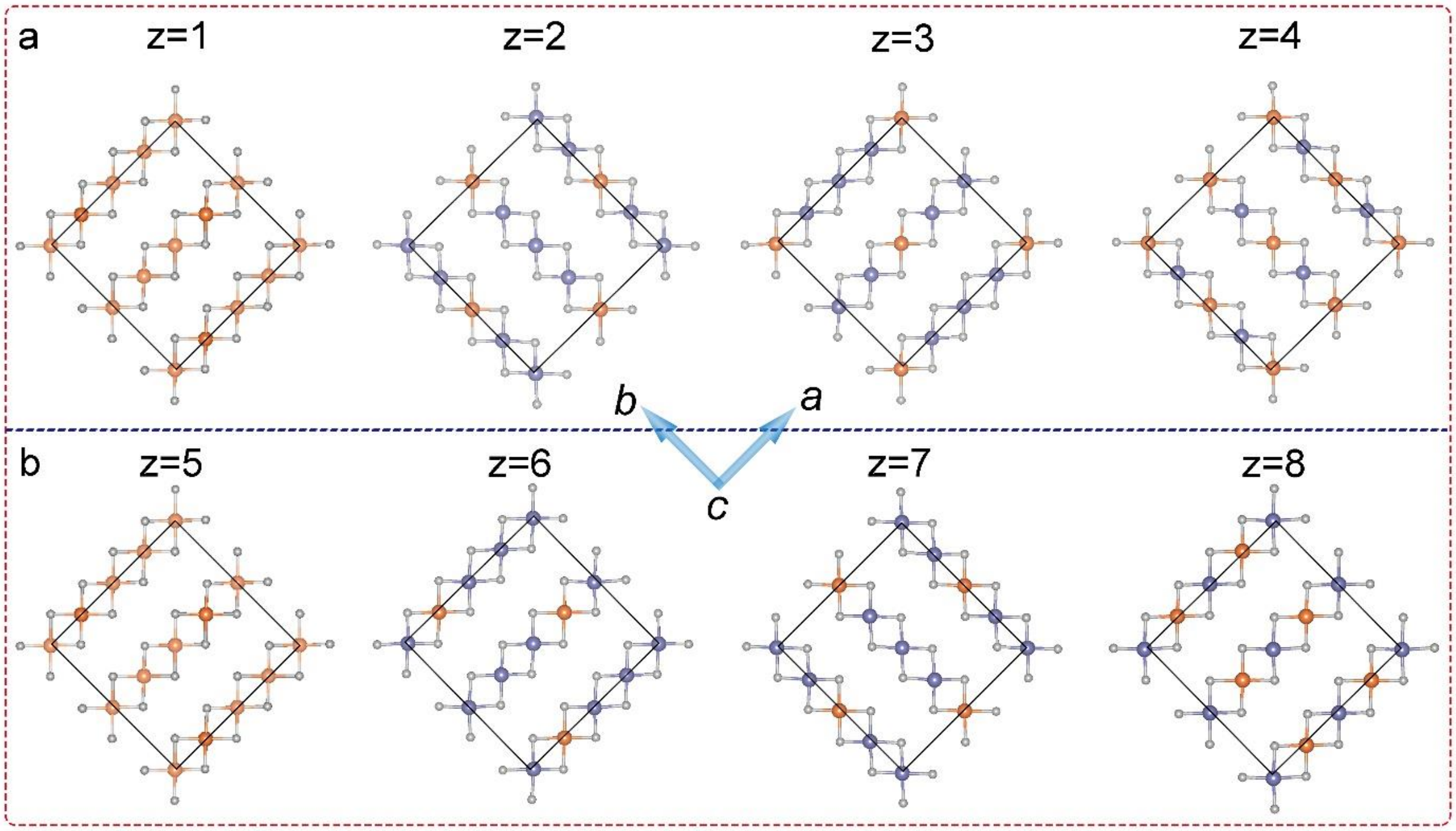

**Fig. S14.**
**Charge order arrangement in each layer in the monoclinic phase.** The measurement result is from in(*2*). In the monoclinic structure, there are eight layers of octahedral Fe ions in one-unit cell. z is the layer number along ***c*** axis. The purple and orange spheres represent Fe ions with different valence states, e.g., $Fe^{3+}$ and $Fe^{2+}$, which are rounded to the nearest integer.

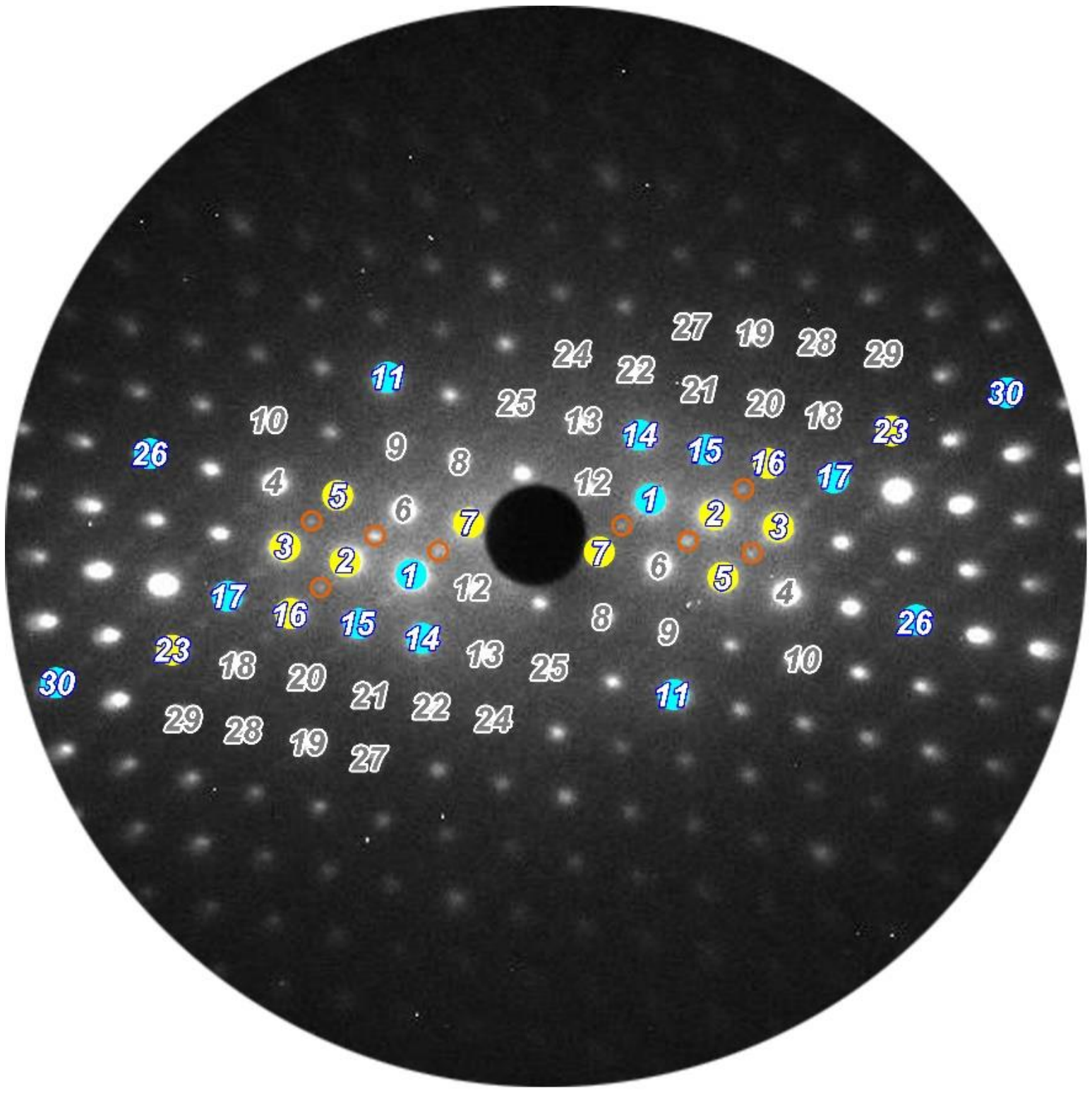

**Fig. S15.**
**UED pattern at 30 K.** The measured Bragg peaks are marked with numbers. The equivalent peaks are using the same number. The Bragg peaks shown with blue and yellow color present clear intensity decreasing and increasing tendency after photoexcitation, respectively. The SLs highlighted by circles.

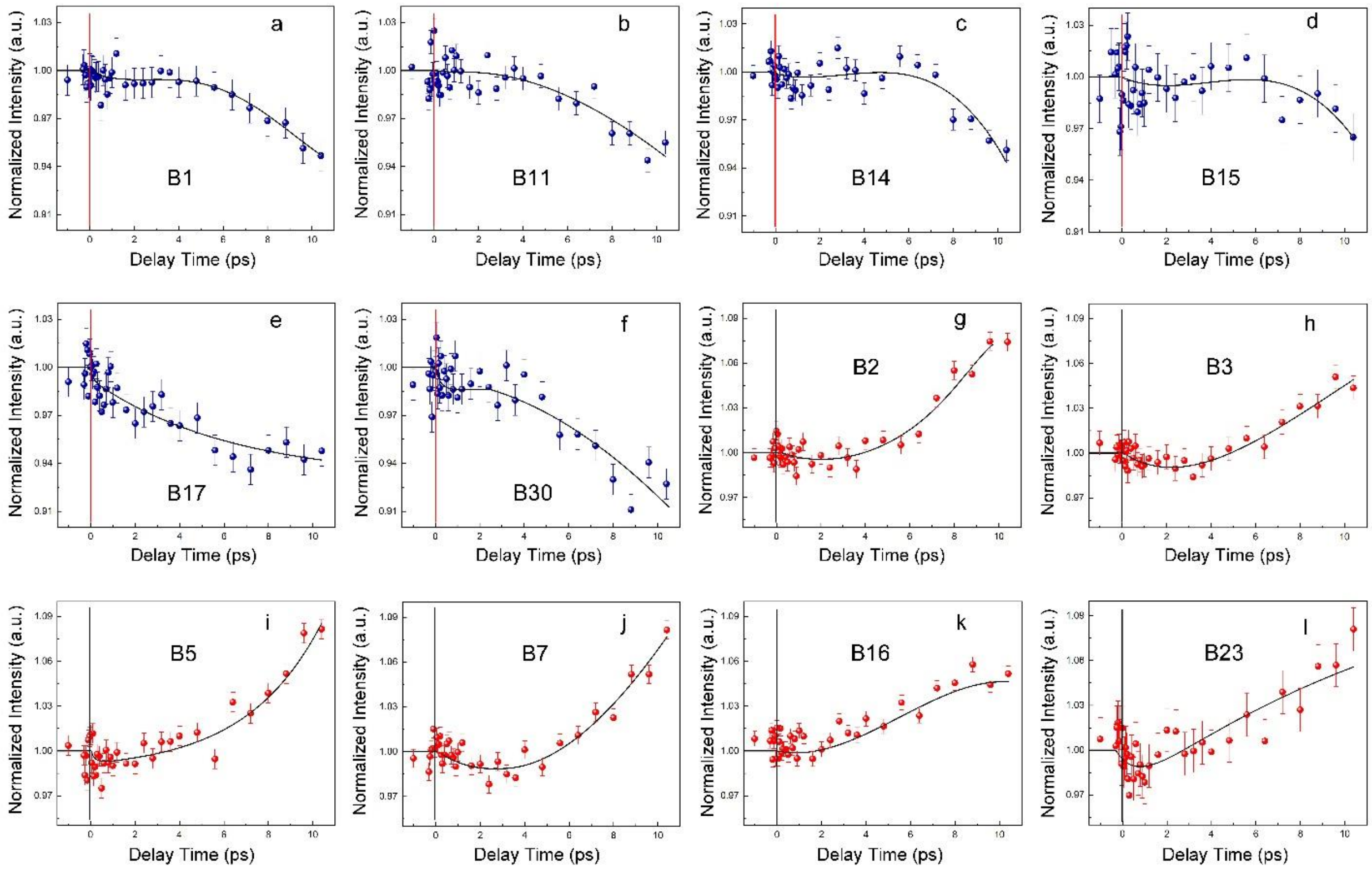

**Fig. S16.**

**Temporal evolutions of the Bragg peaks. a-f** Intensities of Bragg peaks are decreasing; **g-l** Intensities of Bragg peaks are increasing as a function of time. Error bars represent the standard deviation in the mean of intensity before time zero. The solid line is a guide to the eye.

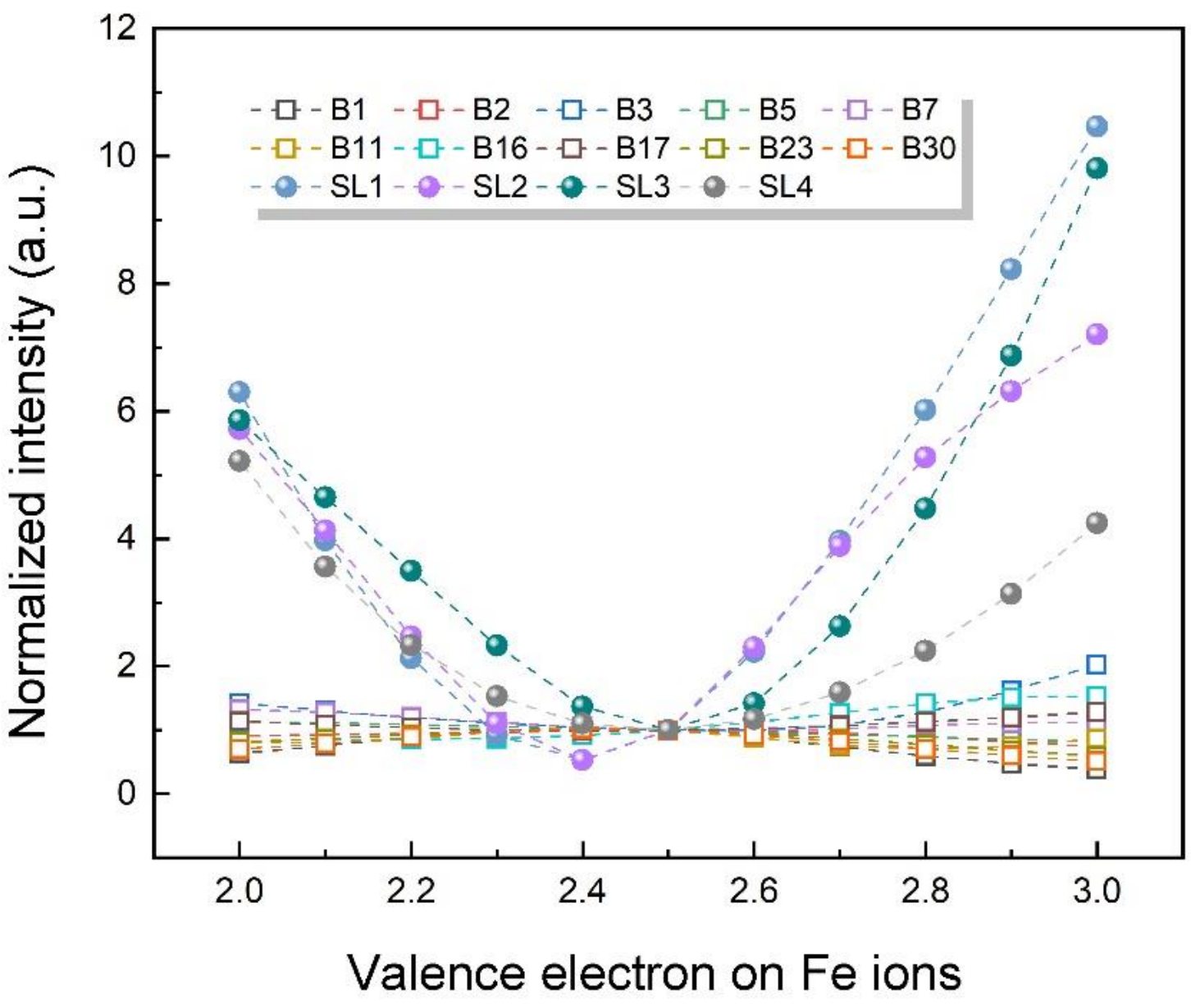

**Fig. S17.**

**Electron diffraction calculation result on charge discrepancy on Fe ions in the monoclinic phase.** The peak intensity is set to 1 as a reference, when the valence electron on the Fe ions is 2.5+, i.e., charge discrepancy is zero. The intensity variations induced by the charge discrepancy on each Bragg peak and SL are calculated as a function of charge disproportion on Fe ions.

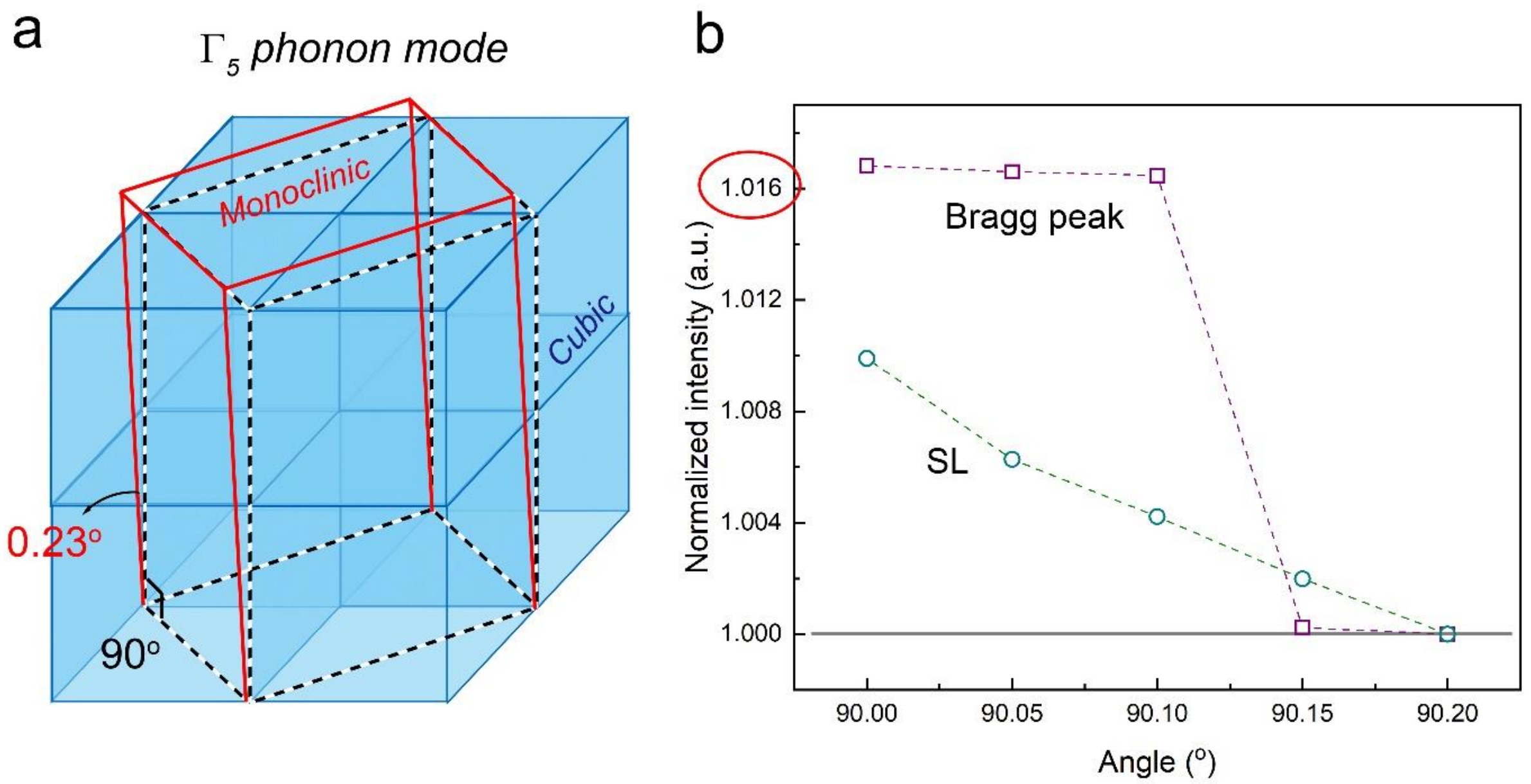

**Fig. S18.**

**$\Gamma_5$ phonon mode. a** Sketch of phonon $\Gamma_5$ mode, which is related to the β angle change during the phase transition from 90˚ in the cubic phase to 90.23˚ in the monoclinic phase. **b** Bragg peak 1 and SL 1 Intensity variation induced by angle β changing. The intensity in the monoclinic phase is the reference as 1.

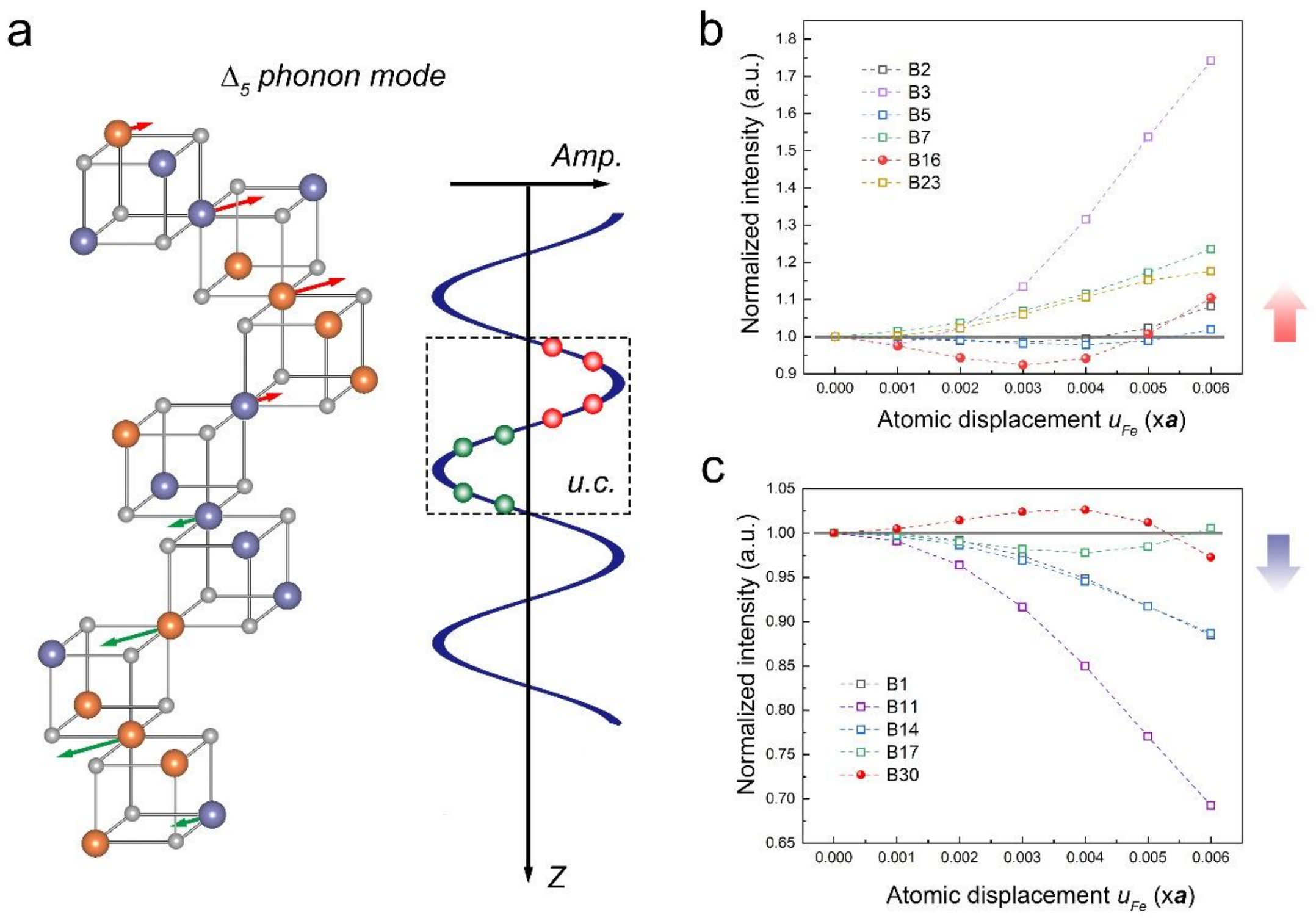

**Fig. S19.**

**$\Delta_5$ phonon mode. a** Atomic displacements model corresponding to the $\Delta_5$ phonon mode. The arrows present the direction of atomic displacements, and the length of arrow is proportional to the displacement amplitude, which is shown in the sine wave curve in the right part. The red dots and green dots show the relative displacement in each layer in one unit cell along ***z*** direction. **b** and **c** Calculated intensities for the Bragg peaks we studied. The red arrow indicates the measured intensity of the Bragg peaks shown in **b** are increasing as a function of time in the experiment; the bule arrow indicate the intensities of those reflection in **c** are decreasing with time in the experimental observation.

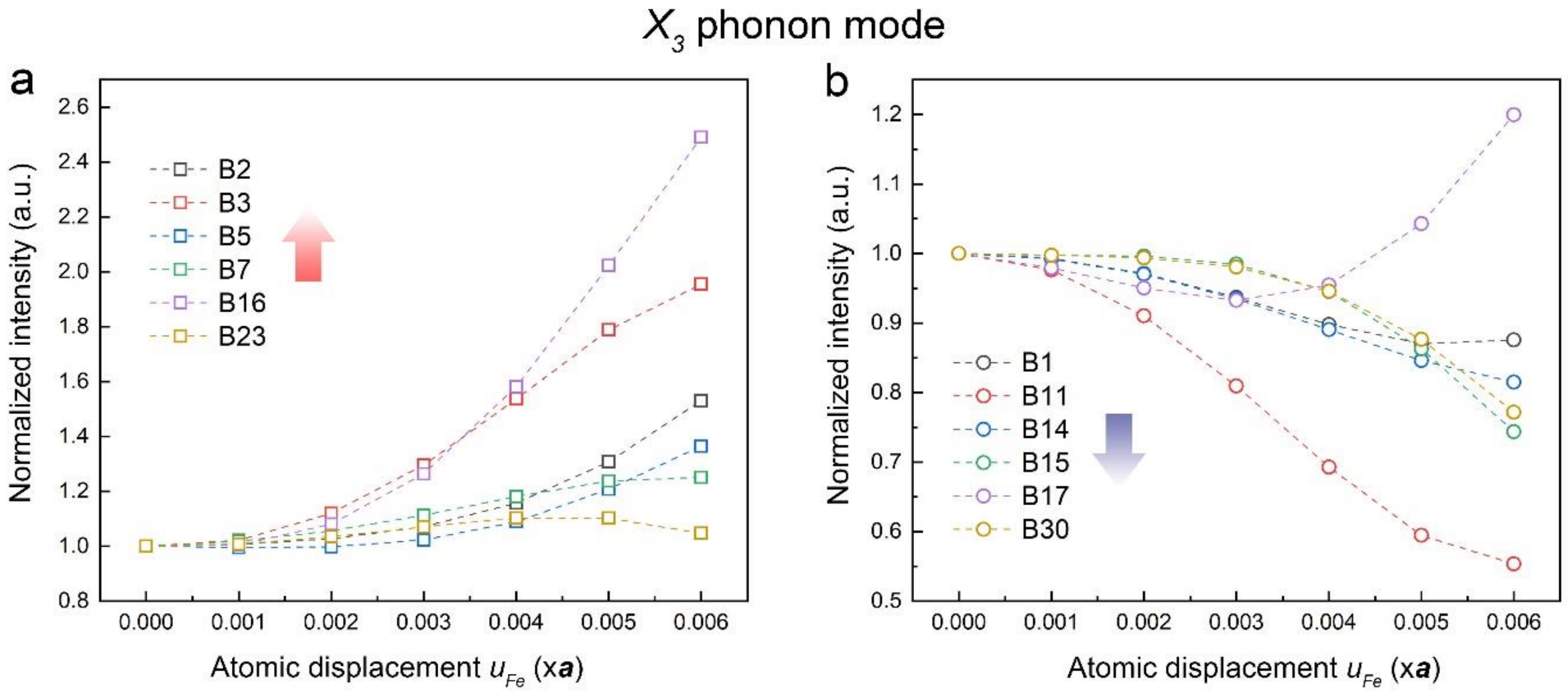

**Fig. S20.**

**Calculation results based on $X_3$ phonon mode. a** and **b** Calculated intensities for the Bragg peaks based on $X_3$ phonon mode. The red arrow indicates the measured intensity of the Bragg peaks shown in **a** are increasing as a function of time; the blue arrow indicates the intensities of those reflection in **b** are decreasing with time measured from the experimental data.

**Table S1. Wyckoff position in $Fd\bar{3}m$ cubic phase**

| Atom | Multiplicity | x | y | z |
|---|---|---|---|---|
| Fe (*tet.*) | 8*a* | 1/8 | 1/8 | 1/8 |
| Fe (*oct.*) | 16*d* | 1/2 | 1/2 | 1/2 |
| O | 32*e* | 0.2549 | 0.2549 | 0.2549 |

**Table S2. Fe Atom position in the octahedral sites**

| F.C.C | | (0, 0, 0) | (0, 1/2, 1/2) | (1/2, 0, 1/2) | (1/2, 1/2, 0) |
|---|---|---|---|---|---|
| *subgroup 1* | $f_1$ | (1/2, 1/2, 1/2) | (1/2, 0, 0) | (0, 1/2, 0) | (0, 0, 1/2) |
| *subgroup 2* | $f_2$ | (1/4, 3/4, 0) | (1/4, 1/4, 1/2) | (3/4, 3/4, 1/2) | (3/4, 1/4, 0) |
| *subgroup 3* | $f_3$ | (3/4, 0, 1/4) | (3/4, 1/2, 3/4) | (1/4, 0, 3/4) | (1/4, 1/2, 1/4) |
| *subgroup 4* | $f_4$ | (0, 1/4, 3/4) | (0, 3/4, 1/4) | (1/2, 1/4, 1/4) | (1/2, 3/4, 3/4) |

Where $f_i$, $i$ =1, 2, 3, 4, is the atom form factor of Fe ions on the 16*d* site.

**Table S3. Atomic displacements ($\delta_i$):**

| F.C.C | (0, 0, 0) | (0, 1/2, 1/2) | (1/2, 0, 1/2) | (1/2, 1/2, 0) |
|---|---|---|---|---|
| *subgroup 1* | (1/2+$\delta_1$, 1/2, 1/2) | (1/2+$\delta_1$, 0, 0) | ($\delta_1$, 1/2, 0) | ($\delta_1$, 0, 1/2) |
| *subgroup 2* | (1/4+$\delta_2$, 3/4, 0) | (1/4+$\delta_2$, 1/4, 1/2) | (3/4+$\delta_2$, 3/4, 1/2) | (3/4+$\delta_2$, 1/4, 0) |
| *subgroup 3* | (3/4+$\delta_2$, 0, 1/4) | (3/4+$\delta_2$, 1/2, 3/4) | (1/4+$\delta_2$, 0, 3/4) | (1/4+$\delta_2$, 1/2, 1/4) |
| *subgroup 4* | ($\delta_2$, 1/4, 3/4) | ($\delta_2$, 3/4, 1/4) | (1/2+$\delta_2$, 1/4, 1/4) | (1/2+$\delta_2$, 3/4, 3/4) |

**Table S4. Bragg peaks included in the diffraction simulation**

| | **B1** | **B2** | **B3** | **B5** | **B7** | **B11** |
|---|---|---|---|---|---|---|
| $[100]_{LT}$ | $(0\ 2\ \bar{6})$ | $(0\ 4\ \bar{8})$ | $(0\ 6\ \overline{10})$ | $(0\ 6\ \bar{6})$ | $(0\ 2\ \bar{2})$ | $(0\ 8\ 0)$ |
| | $(0\ \bar{2}\ 6)$ | $(0\ \bar{4}\ 8)$ | $(0\ \bar{6}\ 10)$ | $(0\ \bar{6}\ 6)$ | $(0\ \bar{2}\ 2)$ | $(0\ \bar{8}\ 0)$ |
| $[010]_{LT}$ | $(2\ 0\ 6)$ | $(4\ 0\ 8)$ | $(6\ 0\ 10)$ | $(6\ 0\ 6)$ | $(2\ 0\ 2)$ | $(8\ 0\ 0)$ |
| | $(\bar{2}\ 0\ \bar{6})$ | $(\bar{4}\ 0\ \bar{8})$ | $(\bar{6}\ 0\ \overline{10})$ | $(\bar{6}\ 0\ \bar{6})$ | $(\bar{2}\ 0\ \bar{2})$ | $(\bar{8}\ 0\ 0)$ |
| $[111]_{LT}$ | $(4\ \bar{2}\ \bar{2})$ | $(6\ \bar{2}\ \bar{4})$ | $(8\ \bar{2}\ \bar{6})$ | $(6\ 0\ \bar{6})$ | $(2\ 0\ \bar{2})$ | $(\bar{4}\ \bar{4}\ 8)$ |
| | $(\bar{4}\ 2\ 2)$ | $(\bar{6}\ 2\ 4)$ | $(\bar{8}\ 2\ 6)$ | $(\bar{6}\ 0\ 6)$ | $(\bar{2}\ 0\ 2)$ | $(4\ 4\ \bar{8})$ |
| $[\bar{1}11]_{LT}$ | $(2\ 4\ \bar{2})$ | $(2\ 6\ \bar{4})$ | $(2\ 8\ \bar{6})$ | $(0\ 6\ \bar{6})$ | $(0\ 2\ \bar{2})$ | $(4\ \bar{4}\ 8)$ |
| | $(\bar{2}\ \bar{4}\ 2)$ | $(2\ 6\ \bar{4})$ | $(\bar{2}\ \bar{8}\ 6)$ | $(0\ \bar{6}\ 6)$ | $(0\ \bar{2}\ 2)$ | $(\bar{4}\ 4\ \bar{8})$ |
| $[1\bar{1}1]_{LT}$ | $(2\ 4\ 2)$ | $(2\ 6\ 4)$ | $(2\ 8\ 6)$ | $(0\ 6\ 6)$ | $(0\ 2\ 2)$ | $(\bar{4}\ 4\ 8)$ |
| | $(\bar{2}\ \bar{4}\ \bar{2})$ | $(\bar{2}\ \bar{6}\ \bar{4})$ | $(\bar{2}\ \bar{8}\ \bar{6})$ | $(0\ \bar{6}\ \bar{6})$ | $(0\ \bar{2}\ \bar{2})$ | $(4\ \bar{4}\ \bar{8})$ |
| $[11\bar{1}]_{LT}$ | $(2\ \bar{4}\ \bar{2})$ | $(2\ \bar{6}\ \bar{4})$ | $(2\ \bar{8}\ \bar{6})$ | $(0\ 6\ 6)$ | $(0\ 2\ 2)$ | $(4\ 4\ 8)$ |
| | $(\bar{2}\ 4\ 2)$ | $(\bar{2}\ 6\ 4)$ | $(\bar{2}\ 8\ 6)$ | $(0\ \bar{6}\ \bar{6})$ | $(0\ \bar{2}\ \bar{2})$ | $(\bar{4}\ \bar{4}\ \bar{8})$ |

| | **B14** | **B15** | **B16** | **B17** | **B23** | **B30** |
|---|---|---|---|---|---|---|
| $[100]_{LT}$ | $(0\ 0\ 8)$ | $(0\ 2\ \overline{10})$ | $(0\ 4\ \overline{12})$ | $(0\ 6\ \overline{14})$ | $(0\ 6\ \overline{18})$ | $(0\ 8\ \overline{20})$ |
| | $(0\ 0\ \bar{8})$ | $(0\ \bar{2}\ 10)$ | $(0\ \bar{4}\ 12)$ | $(0\ \bar{6}\ 14)$ | $(0\ \bar{6}\ 18)$ | $(0\ \bar{8}\ 20)$ |
| $[010]_{LT}$ | $(0\ 0\ 8)$ | $(2\ 0\ 10)$ | $(4\ 0\ 12)$ | $(6\ 0\ 14)$ | $(6\ 0\ 18)$ | $(8\ 0\ 20)$ |
| | $(0\ 0\ \bar{8})$ | $(\bar{2}\ 0\ \overline{10})$ | $(\bar{4}\ 0\ \overline{12})$ | $(\bar{6}\ 0\ \overline{14})$ | $(\bar{6}\ 0\ \overline{18})$ | $(\bar{8}\ 0\ \overline{20})$ |
| $[111]_{LT}$ | $(4\ \bar{4}\ 0)$ | $(\bar{6}\ 4\ 2)$ | $(8\ \bar{4}\ \bar{4})$ | $(10\ \bar{4}\ \bar{6})$ | $(12\ \bar{6}\ \bar{6})$ | $(14\ \bar{6}\ \bar{8})$ |
| | $(\bar{4}\ 4\ 0)$ | $(6\ \bar{4}\ \bar{2})$ | $(\bar{8}\ 4\ 4)$ | $(\overline{10}\ 4\ 6)$ | $(\overline{12}\ 6\ 6)$ | $(\overline{14}\ 6\ 8)$ |
| $[\bar{1}11]_{LT}$ | $(4\ 4\ 0)$ | $(\bar{4}\ \bar{6}\ 2)$ | $(4\ 8\ \bar{4})$ | $(4\ 10\ \bar{6})$ | $(6\ 12\ \bar{6})$ | $(6\ 14\ \bar{8})$ |
| | $(\bar{4}\ \bar{4}\ 0)$ | $(4\ 6\ \bar{2})$ | $(\bar{4}\ \bar{8}\ 4)$ | $(\bar{4}\ \overline{10}\ 6)$ | $(\bar{6}\ \overline{12}\ 6)$ | $(\bar{6}\ \overline{14}\ 8)$ |
| $[1\bar{1}1]_{LT}$ | $(4\ 4\ 0)$ | $(\bar{4}\ \bar{6}\ \bar{2})$ | $(4\ 8\ 4)$ | $(4\ 10\ 6)$ | $(6\ 12\ 6)$ | $(6\ 14\ 8)$ |
| | $(\bar{4}\ \bar{4}\ 0)$ | $(4\ 6\ 2)$ | $(\bar{4}\ \bar{8}\ \bar{4})$ | $(\bar{4}\ \overline{10}\ \bar{6})$ | $(\bar{6}\ \overline{12}\ \bar{6})$ | $(\bar{6}\ \overline{14}\ \bar{8})$ |
| $[11\bar{1}]_{LT}$ | $(4\ \bar{4}\ 0)$ | $(4\ \bar{6}\ \bar{2})$ | $(4\ \bar{8}\ \bar{4})$ | $(4\ \overline{10}\ \bar{6})$ | $(6\ \overline{12}\ \bar{6})$ | $(6\ \overline{14}\ \bar{8})$ |
| | $(\bar{4}\ 4\ 0)$ | $(\bar{4}\ 6\ 2)$ | $(\bar{4}\ 8\ 4)$ | $(\bar{4}\ 10\ 6)$ | $(\bar{6}\ 12\ 6)$ | $(\bar{6}\ 14\ 8)$ |

**Table S5. Superlattice peaks included in the diffraction simulation**:

| | SL1 | SL2 | SL3 | SL4 |
|---|---|---|---|---|
| $[100]_{LT}$ | $(0\ \bar{4}\ 6)$ | $(0\ \bar{2}\ 4)$ | $(0\ \bar{4}\ 10)$ | $(0\ \bar{6}\ 12)$ |
| | $(0\ 4\ \bar{6})$ | $(0\ 2\ \bar{4})$ | $(0\ 4\ \overline{10})$ | $(0\ 6\ \overline{12})$ |
| $[010]_{LT}$ | $(4\ 0\ 6)$ | $(2\ 0\ 4)$ | $(4\ 0\ 10)$ | $(6\ 0\ 12)$ |
| | $(\bar{4}\ 0\ \bar{6})$ | $(\bar{2}\ 0\ \bar{4})$ | $(\bar{4}\ 0\ \overline{10})$ | $(\bar{6}\ 0\ \overline{12})$ |
| $[111]_{LT}$ | $(5\ \bar{1}\ \bar{4})$ | $(3\ \bar{1}\ \bar{2})$ | $(7\ \bar{3}\ \bar{4})$ | $(9\ \bar{3}\ \bar{6})$ |
| | $(\bar{5}\ 1\ 4)$ | $(\bar{3}\ 1\ 2)$ | $(\bar{7}\ 3\ 4)$ | $(\bar{9}\ 3\ 6)$ |
| $[\bar{1}11]_{LT}$ | $(1\ 5\ \bar{4})$ | $(1\ 3\ \bar{2})$ | $(3\ 7\ \bar{4})$ | $(3\ 9\ \bar{6})$ |
| | $(\bar{1}\ \bar{5}\ 4)$ | $(\bar{1}\ \bar{3}\ 2)$ | $(\bar{3}\ \bar{7}\ 4)$ | $(\bar{3}\ \bar{9}\ 6)$ |
| $[1\bar{1}1]_{LT}$ | $(1\ 5\ 4)$ | $(1\ 3\ 2)$ | $(3\ 7\ 4)$ | $(3\ 9\ 6)$ |
| | $(\bar{1}\ \bar{5}\ \bar{4})$ | $(\bar{1}\ \bar{3}\ \bar{2})$ | $(\bar{3}\ \bar{7}\ \bar{4})$ | $(\bar{3}\ \bar{9}\ \bar{6})$ |
| $[11\bar{1}]_{LT}$ | $(1\ \bar{5}\ \bar{4})$ | $(1\ \bar{3}\ \bar{2})$ | $(3\ \bar{7}\ \bar{4})$ | $(3\ \bar{9}\ \bar{6})$ |
| | $(\bar{1}\ 5\ 4)$ | $(\bar{1}\ 3\ 2)$ | $(\bar{3}\ 7\ 4)$ | $(\bar{3}\ 9\ 6)$ |